\documentclass[pra, twocolumn,letterpaper, showpacs]{revtex4}

\usepackage{amsmath}
\usepackage{verbatim}
\usepackage{amsfonts}
\usepackage{amssymb}
\usepackage{graphicx}
\usepackage{color}

\begin{document}
\begin{abstract}
We theoretically investigate interactions between non-degenerate mechanical oscillators mediated by a time-dependent cavity field. We obtain a reduced master equation valid for all optomechanical systems operating in the weak coupling regime. This master equation includes all forms of decoherence and back-action due to the dissipation of the field mediating the interaction. We apply the master equation to study two resonant coupling schemes within a rotating-wave approximation: the beam splitter Hamiltonian and the two-mode parametric amplifier. In both cases, the effective unitary interaction can be made arbitrarily strong compared to the decoherence due to dissipation of the mediating field by choosing appropriate detunings. 
\end{abstract}
\title{Non-Degenerate Multimode Optomechanics}
\author{L.~F.~Buchmann}
\affiliation{Department of Physics, University of California, Berkeley, California 94720, USA}
\author{D.~M.~Stamper-Kurn}
\affiliation{Department of Physics, University of California, Berkeley, California 94720, USA}
\affiliation{Materials Sciences Division, Lawrence Berkeley National Laboratory, Berkeley, CA 94720, USA}
\maketitle
\section{Introduction}
The ability to control a physical system at the quantum level has, whenever it was achieved, profoundly influenced physics and many of its applications. Two prominent examples, the development of lasers and their application to cool atomic gases, have allowed to use light as a tool to perform a wide range of tasks and lead to improvements of precision measurements by orders of magnitude. In the last few years, successes in cavity optomechanics have lead to similar control over motional degrees of freedom of mesoscopic and microscopic oscillators~\cite{review}.
\par
Experiments have successfully demonstrated quantum effects involving a single mechanical oscillator~\cite{OMsqueezing, Cleland, ultracold, sidebandasym, forcesensitivity, palomaki} and theoretical studies point out the suitability of optomechanical systems to observe truly macroscopic non-classical behavior~\cite{nonclassicalOM, MAQRO, tunneling,nonclassicalOM2, Yanbei}. While the vast majority of the experimental results involve a single mechanical mode, typical solid-state resonators exhibit a multitude of resonances over a wide range of frequencies~\cite{regalmultimode, mukundmultimode}. In systems where the mechanical modes consist of the collective motion of a trapped atomic cloud~\cite{ultracold, Esslinger} one can engineer non-degenerate optomechanical modes by trapping several clouds in an optical superlattice~\cite{coldatomsmultimode}. Numerous non-degenerate mechanical modes are thus a common feature of optomechanical systems, which feature, however, remains largely unexploited to date as most experimental results~\cite{multimodeOM5, multimodeOM6, multimodeOM3, regalmultimode} involve nearly degenerate mechanical modes.
\par
While multimode optomechanics has been studied theoretically for specific applications~\cite{mattash, multimodeOM1, multimodeOM2, multimodeOM3, multimodeOM4, phasecon, mediatedinteractions, borkje}, there has been no general formalism to analyze the possibilities of tuning non-degenerate modes into resonance. The aim of the present work is to provide that formalism, including the effects of quantum noise and measurement back-action of the mediating field. The resulting equations are applicable over the whole range of optomechanical systems, from large scale interferometers for gravitational wave detection~\cite{gravwave} to oscillations of ultracold atomic clouds~\cite{ultracold} or optomechanical microcavities~\cite{review,microcavity}.\par
The effective interaction between non-degenerate mechanical modes is suppressed by their frequency difference. Together with the typically small optomechanical couplings this prohibits effective multimode physics involving modes with vastly different frequencies. However, other fields of physics successfully use auxiliary modes to resonantly couple non-degenerate states. 
In atomic physics, for instance, two laser beams appropriately detuned from a third state can  drive a two-photon resonance between two non-degenerate atomic states~\cite{textbook}. In non-linear optics, two optical fields can be tuned into resonance by an appropriate pump creating a standing wave in a non-linear crystal~\cite{textbook}. The optomechanical analogue of these techniques is an oscillating intensity in a cavity field coupled to two non-degenerate mechanical modes and is the subject of this Article. \par
A time dependent intracavity intensity is created by any anharmonic pump, the simplest example being the field created by two harmonic drives with distinct frequencies. The resulting cavity field can mediate a resonant coupling irrespective of the frequency difference of the mechanical modes. Recent theoretical studies have applied specific examples of this very effect for specialized applications, such as the cooling of a low-lying mechanical mode~\cite{borkje}, mechanical phase-conjugation~\cite{phasecon} or preparation of non-classical states~\cite{mattash}. The price for resonant mediated interaction between non-degenerate modes is additional decoherence due to the dissipative nature of the mediating field. The cavity field that provides mechanical coupling also carries information about the mechanical state out of the cavity and this loss of unitarity is necessarily accompanied by noise entering the mechanical system~\cite{mediatedinteractions}. \par
In this Article, we derive a general and complete master equation for the behavior of two non-degenerate mechanical modes interacting via weak coupling to a shared cavity field. Our treatment is complete in the sense that  all sources of decoherence, back-action and noise that are a consequence of the mediating field's dissipation are included. Taking these effects into account is crucial for the operation of any optomechanical device at the quantum level and a prerequisite to the observation of non-classical effects. The resulting equation is general in the sense that we do not assume any specific detuning or rely on the system's being in any particular parameter regime, apart from the weak coupling assumption which is accessible to virtually all optomechanical systems to date~\cite{review}. The derivation of our master equation does not rely on the mediating field's being the mode of an electromagnetic resonator. With a few simple modifications it will also describe interactions mediated by bulk-phononic modes~\cite{mukundmed}. Similarly it is straightforward to add individual, Markovian heat baths for the mechanical oscillators and while we have neglected them for the discussion of the individual couplings, their effect is discussed in the final section. 
\par
After deriving the general master equation, we will study the two resonant effective interactions described by the beam splitter Hamiltonian
\begin{equation}\label{BShamiltonian}
\mathcal{H}_\mathrm{I}=J_\mathrm{BS}\left(\hat{b}_1^\dag\hat{b}_2+\hat{b}_1\hat{b}_2^\dag\right)
\end{equation}
and the two-mode parametric amplifier
\begin{equation}\label{PAhamiltonian}
\mathcal{H}_\mathrm{I}=J_\mathrm{PA}\left(\hat{b}_1\hat{b}_2+\hat{b}_1^\dag\hat{b}_2^\dag\right),
\end{equation}
where $\hat{b}_j$ denotes the annihilation operator for phonons in oscillator $j$. The former allows the mapping of a quantum state from one oscillator to the other while the latter leads to non-classical correlations between the two oscillators, akin to parametric down-conversion in quantum optics. Most results obtained in single-mode optomechanics consist of these two Hamiltonians resonantly coupling the cavity-mode with the mechanical mode via the matching of mechanical side bands with the cavity resonance. Optomechanical cooling, for instance, is a realization of the beam splitter Hamiltonian together with the large dissipation of the electromagnetic field modes, while parametric driving is a realization of the parametric amplifier involving mechanical and electromagnetic modes~\cite{palomaki}. Realizing these couplings between two mechanical modes opens the door to more customized and precise sensing of forces and fields, the application of quantum computation algorithms to mechanical systems~\cite{computing} and the study of macroscopic decoherence~\cite{diosi}. \par
Of crucial importance is the ratio between the strength of the mediated unitary interactions and the decoherence rates due to the mediating field being emitted from the cavity. This quantity is a measure for the ability of the effective interaction to create or maintain non-classical correlations between the two modes and thus a gauge for the classicality of the interactions~\cite{mediatedinteractions, kafri1, kafri2}. We find that the relationship between unitary coupling strengths and total decoherence rates depends on the system parameters and the detunings of the pumps with respect to the cavity resonance, such that different choices of detunings turn out to be ideal for a given application. Most importantly, we find that even for systems where the cavity emission rate of the mediating field exceeds all of the involved mechanical frequencies there are regimes where the engineered interaction allows non-classical motional states or the transfer of nonclassical states between oscillators. 

\section{Master Equation}
\label{mastereqsection}
\subsection{System Hamiltonian}
\begin{figure}
\includegraphics[width=1\columnwidth]{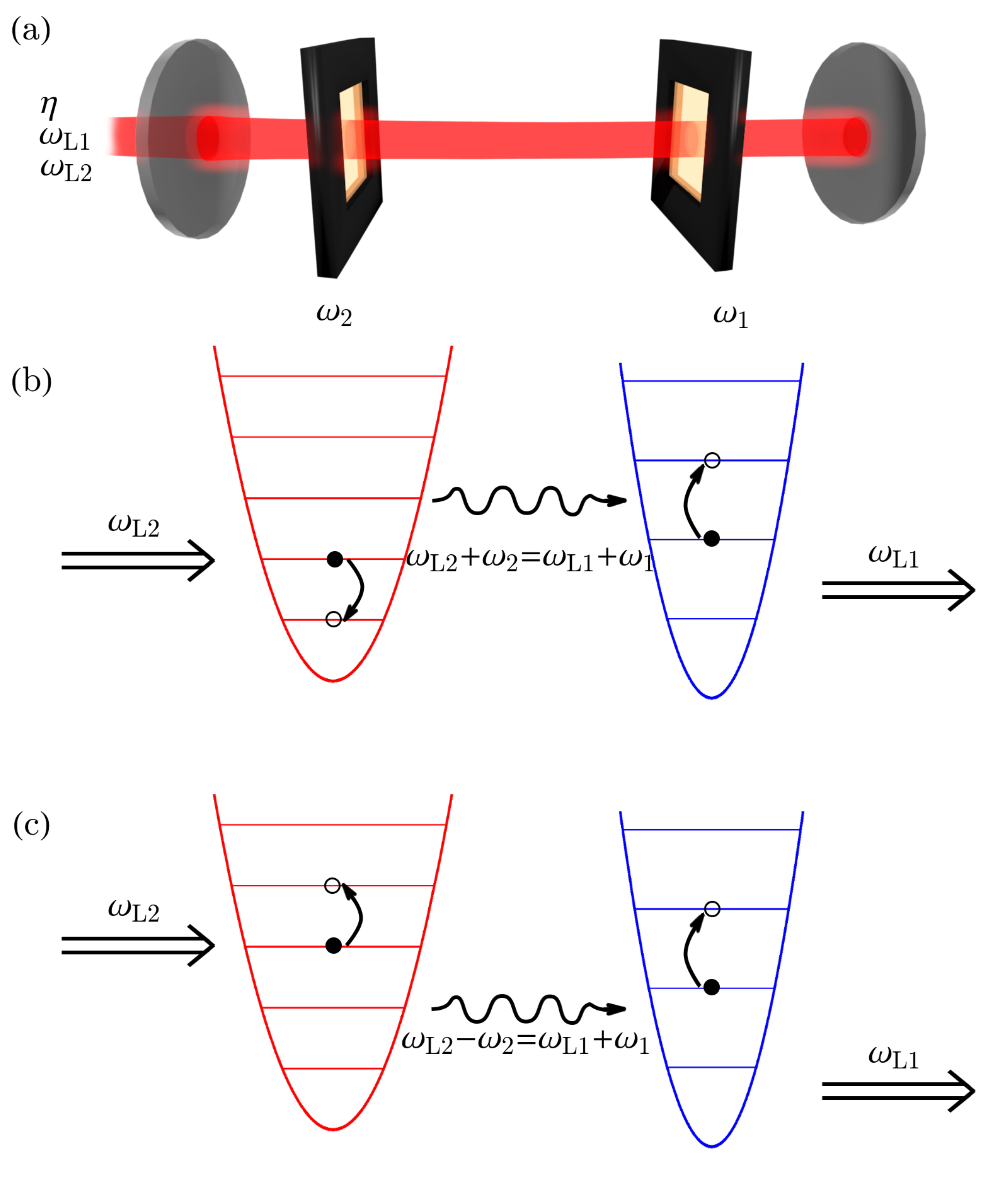}
\caption{(Color online) (a) Sketch of a possible experimental realization involving separate optomechanical membranes. The mechanical modes could also be on the same resonator.  (b) A scattering process contributing to the coherent exchange of phonons in the resonant beam splitter interaction. The three other contributing processes result from time reversal and pump photons with frequency $\omega_{L1}$ scattering with oscillator 2. (c) A scattering process contributing to the parametric amplification of the mechanical modes. The three other processes are obtained in the same way as in panel (b).}
\label{setupsketch}
\end{figure}

Consider an optomechanical system consisting of two mechanical modes with natural frequencies $\omega_1$ and $\omega_2$ coupled to the field of an electromagnetic resonator with single photon/single phonon coupling strengths $g_1$ and $g_2$ respectively.A sketch of a possible setup involving spatially separate membranes coupled to the field of an optical Fabry-P\'erot resonator is given in Fig.~\ref{setupsketch}. The mechanical modes need not be located on distinct oscillators and the mediating field does not have to be optical. The system could also consist of two non-degenerate drumhead modes of a capacitor in a superconducting LC circuit~\cite{palomaki} with photons in the microwave regime or two samples of ultracold atoms trapped in an optical cavity~\cite{coldatomsmultimode}. 
\par
Neglecting all other modes of the mechanical system and the cavity, the Hamiltonian in the lab frame is given by 
\begin{equation}\label{labframehamiltonian}
\mathcal{H}=\mathcal{H}_\mathrm{opt}+\mathcal{H}_\mathrm{m}+\mathcal{H}_\mathrm{I}+\mathcal{H}_\kappa
\end{equation}
with 
\begin{align}
\mathcal{H}_\mathrm{opt}&=\omega_c\hat{a}^\dag\hat{a}+\eta(t)\hat{a}^\dag+\eta^*(t)\hat{a},
\end{align}
\begin{align}
\mathcal{H}_\mathrm{m}&=\sum_{j=1}^2\omega_j\hat{b}_j^\dag\hat{b}_j,
\end{align}
\begin{align}
\mathcal{H}_\mathrm{I}&=\sum_{j=1}^2g_j\hat{a}^\dag\hat{a}(\hat{b}_j^\dag+\hat{b}_j),
\end{align}
and $\mathcal{H}_\kappa$ accounts for the intensity of the intracavity field being emitted to a reservoir at zero temperature at a rate $\kappa$  \footnote{The frequencies of electromechanical systems working in the microwave regime cannot be assumed to feature a vanishing thermal photon occupation. However, typical devices are operated cryogenically, where a zero temperature reservoir is reasonable.}. We have set $\hbar=1$, dropped constant terms and defined the input coherent field driving the cavity with the rate $\eta$. The coupling of each oscillator to its thermal reservoir is neglected in favor of a focus on decoherence due to the dissipative nature of the field that mediates the interaction. Thermal effects are explicitly discussed later. 
\par
To tune a desired coupling into resonance, we need the cavity intensity to oscillate in time. This is achieved by driving the cavity with a bi-chromatic pump,
\begin{equation}
\eta(t)=\eta_1e^{-i\omega_{L1}t}+\eta_2e^{-i\omega_{L2}t}.
\end{equation}
The beat-note between the two drives will provide the necessary energy to tune either Hamiltonian (\ref{BShamiltonian}) or (\ref{PAhamiltonian}) into resonance. 
The density operator of the system $\rho$ satisfies the master equation
\begin{equation}\label{mastereq1}
\frac{d\rho}{dt}=-i[\mathcal{H},\rho]+\kappa\mathcal{L}(\hat{a})\rho,
\end{equation}
with the Lindblad superoperator
\begin{equation}
\mathcal{L}(\hat{a})\rho=\hat{a}\rho\hat{a}^\dag-\frac{1}{2}\hat{a}^\dag\hat{a}\rho-\frac{1}{2}\rho\hat{a}^\dag\hat{a}.
\end{equation}
Without loss of generality we will assume $\omega_1>\omega_2$ and $\omega_{L2}>\omega_{L1}$. The next step is to unitarily displace the cavity field operator by a time-dependent complex function: $\hat{a}\to\alpha(t)+\hat{a}$. If we choose 
\begin{equation}
\alpha(t)=\alpha_1e^{-i\omega_{L1}t}+\alpha_2e^{-i\omega_{L2}t},
\end{equation}
with 
\begin{equation}
\alpha_j=\frac{-i\eta_j}{\kappa/2+i\Delta_j},
\end{equation}
where $\Delta_j=\omega_c-\omega_{Lj}$, all source terms for $\hat{a}$ cancel and it only carries quantum fluctuations of the cavity field. Without loss of generality, we may choose the phases of $\eta_j$ such that both $\alpha_j$ are real. The displacement operation introduces the term $|\alpha(t)|^2\sum_jg_j(\hat{b}_j+\hat{b}_j^\dag)$ driving the mechanical degrees of freedom. The time independent part of this term is a constant force that will shift the equilibrium position of each oscillator. Additionally, the beat-note between the two pumps will drive each oscillator off-resonantly. The constant force does not affect the dynamics of the system. For the small mechanical linewidths typical of optomechanics~\cite{review} the off-resonant terms drive oscillator $j$ into a coherent state with phonon number of the order of $\frac{\alpha_1\alpha_2g_j}{|\omega_{L1}-\omega_{L2}-\omega_j|}$. For weak couplings $g_j$ his small oscillation can be neglected for all effects that are considered in the remainder of this paper. 

Similarly, we can also neglect the interactions quadratic in $\hat{a}$ and thus find in a co-rotating frame for the optical operators the Hamiltonian
\begin{equation}\label{hamiltonian_2}
\mathcal{H}=\mathcal{H}_\mathrm{m}+\sum_{j,k=1}^2g_j\alpha_k\left(\hat{a}^\dag e^{i\Delta_k t}+\hat{a}e^{-i\Delta_kt}\right)(\hat{b}_j+\hat{b}_j^\dag).
\end{equation}

\subsection{Tracing out the cavity field}
Tracing out the electromagnetic system will yield an effective master equation coupling the mechanical modes. 
To perform the partial trace over electromagnetic degrees of freedom we restrict the optical Hilbert space to the optical vacuum and first-order coherences. This is formally a projection of the optical Hilbert space unto a lower-dimensional subspace as described below. Introducing Fock states of the optical subsystem $|n\rangle_\mathrm{o}$ and operators
\begin{equation}
\rho_{mn}={}_\mathrm{o}\langle m | \rho | n \rangle_\mathrm{o},
\end{equation}
the reduction is formalized as
\begin{equation}
\rho_{mn}=0 \iff m+n>1.
\end{equation}
By virtue of the displacement operation this is a reasonable approximation.
The reduced density operator of the mechanical system becomes
\begin{equation}
\rho_\mathrm{m}={}_\mathrm{o}\langle 0 | \rho | 0 \rangle_\mathrm{o}.
\end{equation}
Note, however, that this identification can only be made after taking the trace over optical degrees of freedom in Eq.~(\ref{mastereq1}), as there are non-vanishing contributions from the one-photon sector. Taking this into consideration, we find the equation of motion 
\begin{align}\label{mastereqred}
\frac{d\rho_\mathrm{m}}{dt}=&-i[\mathcal{H}_\mathrm{m},\rho_\mathrm{m}]\nonumber\\
&+\sum_{j,k=1}^2\left(-ig_j\alpha_ke^{-i\Delta_kt}[\hat{b}_j+\hat{b}_j^\dag,\rho_{10}]+h.c.\right),
\end{align}
with $h.c.$ denoting the Hermitian conjugate. 
The evolution of the operator $\rho_{10}$ is governed by
\begin{equation}
\frac{d\rho_{10}}{dt}=-i[\mathcal{H}_m,\rho_{10}]-i\sum_{j,k=1}^2g_j\alpha_ke^{i\Delta_kt}(\hat{b}_j+\hat{b}_j^\dag)\rho_\mathrm{m}-\frac{\kappa}{2}\rho_{10}.
\end{equation}
Assuming that $\rho_\mathrm{m}$ changes slowly on a timescale $\kappa^{-1}$ we can solve this equation by elementary transformations. The result is
\begin{align}
\rho_{10}=&-i\sum_{j,k=1}^2g_j\alpha_ke^{i\Delta_kt}\times\nonumber\\
&\times\left(\frac{\hat{b}_j^\dag}{\kappa/2+i(\Delta_k+\omega_j)}+\frac{\hat{b}_j}{\kappa/2+i(\Delta_k-\omega_j)}\right)\rho_\mathrm{m}.\label{coherences}
\end{align}
This expression can be plugged into Eq.~(\ref{mastereqred}) to give the reduced master equation for the mechanical system. Introducing the vectors $\hat{\textbf{b}}^\top=(\hat{b}_1,\hat{b}_2,\hat{b}_1^\dag,\hat{b}_2^\dag)$, where ${}^\top$ denotes the transpose, we obtain
\begin{equation}\label{mastereqfull}
\frac{d\rho_\mathrm{m}}{dt}=-i[\mathcal{H}_\mathrm{m},\rho_\mathrm{m}]+\hat{\textbf{b}}^\top\mathrm{M}_1\hat{\textbf{b}}\rho_\mathrm{m}+\rho_\mathrm{m}\hat{\textbf{b}}^\top\mathrm{M}_2\hat{\textbf{b}}+\hat{\textbf{b}}^\top\mathrm{M}_3\rho_\mathrm{m}\hat{\textbf{b}}.
\end{equation}
Here $\mathrm{M}_i$ are $4\times 4$ matrices
\begin{align}
\mathrm{M}_i=\left(\begin{array}{c c}
D_i^{-} & O_i^{-}\\
O_i^{+} & D_i^{+}
\end{array}\right),
\end{align}
with
\begin{subequations}\label{matrixelements}
\begin{align}
\left[D_1^\pm\right]_{mn}=-g_mg_n\sum_{k,l=1}^2\frac{\alpha_{k}\alpha_{l}e^{i(\Delta_{l}-\Delta_{k})t}}{\kappa/2+i(\Delta_{l}\pm\omega_n)}\\
\left[O_1^\pm\right]_{mn}=-g_mg_n\sum_{k,l=1}^2\frac{\alpha_{k}\alpha_{l}e^{i(\Delta_{l}-\Delta_{k})t}}{\kappa/2+i(\Delta_{l}\mp\omega_n)}\\
\left[D_2^\pm\right]_{mn}=-g_mg_n\sum_{k,l=1}^2\frac{\alpha_{k}\alpha_{l}e^{i(\Delta_{k}-\Delta_{l})t}}{\kappa/2-i(\Delta_{l}\mp\omega_m)}\\
\left[O_2^\pm\right]_{mn}=-g_mg_n\sum_{k,l=1}^2\frac{\alpha_{k}\alpha_{l}e^{i(\Delta_{k}-\Delta_{l})t}}{\kappa/2-i(\Delta_{l}\mp\omega_m)}
\end{align}
\begin{align}
\left[D_3^\pm\right]_{mn}=&g_mg_n\sum_{k,l=1}^2\alpha_{k}\alpha_{l}\times\nonumber\\
&\times\left(\frac{e^{i(\Delta_{l}-\Delta_{k})t}}{\kappa/2+i(\Delta_{l}\pm\omega_m)}+\frac{e^{i(\Delta_{k}-\Delta_{l})t}}{\kappa/2-i(\Delta_{l}\mp\omega_n)}\right)\\
\left[O_3^\pm\right]_{mn}=&g_mg_n\sum_{k,l=1}^2\alpha_{k}\alpha_{l}\times\nonumber\\
&\times\left(\frac{e^{i(\Delta_{l}-\Delta_{k})t}}{\kappa/2+i(\Delta_{l}\pm\omega_m)}+\frac{e^{i(\Delta_{k}-\Delta_{l})t}}{\kappa/2-i(\Delta_{l}\pm\omega_n)}\right),
\end{align}
\end{subequations}
and $m,n\in\{1,2\}$. 
\par
This is the general form of the master equation in the lab frame, including all  off-resonant terms. 
It is easily reduced to the system of a single harmonic oscillator (with $\omega_1=\omega_2$ and $g_1=g_2$ or $g_2=0$) or a single harmonic pump (with $\Delta_1=\Delta_2$ and $\alpha_1=\alpha_2$ or $\alpha_2=0$). As sketched in Fig.~\ref{setupsketch} (b) and (c), mediated interactions involve at least two photon-phonon scattering processes. Thus all effective amplitudes appearing in Eqs.~(\ref{matrixelements}) are of second order in the single phonon/single photon coupling strength.
\par
The unitary evolution described by Eq. (\ref{mastereqfull}) is conceptually simple. Each pump shifts the resonance of the oscillators by the frequency
\begin{align}\label{lightshift}
\mathrm{Im}\left([O_1^+]_{ii}-[O_2^-]_{ii}\right)\equiv\delta\Omega_i+R_i(t),
\end{align}
where we have separated the time-independent spring shifts $\delta\Omega_i$ from time-dependent contributions easily found from Eqs.~(\ref{matrixelements}). 
The explicit form of the spring shifts is 
\begin{align}
\delta\Omega_i=g_i^2\sum_{k=1}^2\left(\frac{\alpha_k^2(\Delta_k-\omega_i)}{\kappa^2/4+(\Delta_k-\omega_i)^2}-\frac{\alpha_k^2(\Delta_k+\omega_i)}{\kappa^2/4+(\Delta_k+\omega_i)^2}\right).
\end{align}
The mechanical resonances oscillate with small amplitudes $|R_i(t)|$ around their means $\omega_i+\delta\Omega_i$.  Due to the weakness of optomechanical couplings we have $|R_i(t)|,\delta\Omega_i\ll\omega_i$ and the oscillation of their resonance frequency will not affect the dynamics of the system for sufficiently non-degenerate modes, quantified by \mbox{$|\Delta_1-\Delta_2|\gg|\delta\Omega_i|$}. 
Additionally, the light-field also establishes an optical spring between the oscillators, with spring constant given by
\begin{align}\label{springstrength}
k_{12}=\frac{1}{z_\mathrm{HO,1}z_\mathrm{HO,2}}\mathrm{Im}\left([O_1^+]_{12}-[O_2^-]_{21}\right),
\end{align}
where $z_\mathrm{HO,i}$ is the zero-point displacement of oscillator $i$. The re-scaling by the two oscillator lengths serves to give $k_{12}$ the units of force per length but does not appear in the equations of motion, where this length has been absorbed into the definition of the operators $\hat{b}_i$. 
This spring too consists of constant and time-dependent contributions. In contrast to the single-mode optical springs given in Eq. (\ref{lightshift}), it is now the time-dependent contributions that can create resonant couplings between the non-degenerate oscillators.

\section{Resonant Couplings}

In the following section we will focus on the cases where the time-dependence of the optical spring between the oscillator resonates with eigenmodes of the two-oscillator system. We will focus on a regime where the frequency difference between the two oscillators is large compared to the effective coupling strengths,
\begin{align}
\frac{g_1g_2\alpha_1\alpha_2}{\kappa}\ll|\omega_1-\omega_2|\label{nondegeneracy}.
\end{align} 
For such systems, a monochromatic pump does not lead to multimode behavior and the system can be treated as two separate single-mode optomechanical oscillators. In our case of two harmonic pumps, the beat-note between the two drives can tune a desired interaction into resonance. For such cases condition (\ref{nondegeneracy}) assures off-resonant contributions to be suppressed such that they can be neglected within a rotating-wave approximation.
\par
The beat-note between the two optical pumps has to hit the resonances of the optically shifted mechanical frequencies $\omega_i+\delta\Omega_i$. This turns the resonance conditions into non-linear equations with respect to the optical detunings. The correction of the modified resonance conditions compared to their bare values are small, but we have computed them for completeness and they are given in the appendix. From here on, we assume that these corrections are taken into account and take them into the definitions of the mechanical frequencies, i.e. 
\begin{align}
\omega_i\to\omega_i+\delta\Omega_i.
\end{align}
This substitution has to be carried through in Eqs. (\ref{matrixelements}) to be self consistent and does not lead to qualitatively different dynamics in the weak-coupling regime. 

\par
Resonant coupling arises when the modulation frequency of the spring connecting the two oscillators matches the difference of the two individual oscillator resonances, 
\begin{equation}\label{resonancecondition1}
\Delta_1-\Delta_2=\delta\omega
\end{equation}
where we have introduced $\delta\omega=\omega_1-\omega_2$ for the difference of the mechanical frequencies. The resulting interaction -- governed by Hamiltonian (\ref{BShamiltonian}) -- can be used to exchange the quantum states of the two mechanical modes irrespective of their frequency difference. A diagram for one particular set of scattering processes contributing to this interaction is given in Fig.~\ref{setupsketch}~(b). The two classical pumps driving the cavity establish the conditions for resonant exchange  of phonons between the mechanical modes via scattering of intracavity photons. 
Alternatively, the modulation of the coupling spring $k_{12}$ can match the sum of the oscillator frequencies,
\begin{equation}\label{resonancecondition2}
\Delta_1-\Delta_2=2\bar\Omega
\end{equation}
with $\bar\Omega=\frac{\omega_1+\omega_2}{2}$ being the average frequency of the bare mechanical modes.
In this case the resulting interaction is a parametric drive of a collective mode of the two-oscillator system described by Hamiltonian (\ref{PAhamiltonian}). An example for a process contributing coherently to this effective interaction is given in Fig.~\ref{setupsketch}~(c). 

In the following, we will study the two resonant cases in detail. To this end, it is useful to transform into a co-rotating frame for the mechanical operators, achieved by applying the unitary 
\begin{equation}
U=e^{i\sum_j\omega_j\hat{b}_j^\dag\hat{b}_jt}
\end{equation}
to the master equation (\ref{mastereqfull}). The transformed Hamiltonian does not contain the free mechanical Hamiltonians $\mathcal{H}_\mathrm{m}$, but the frequencies now modulate the off-diagonal elements of the matrices appearing in the master equation,
\begin{align}
[D_j^{\pm}]_{mn}&\to e^{\pm i (\omega_m+\omega_n)t}[D_j^{\pm}]_{mn},\\
[O_j^{\pm}]_{mn}&\to e^{\pm i (\omega_m-\omega_n)t}[O_j^{\pm}]_{mn}.
\end{align}

\subsection{Beam Splitter}
\begin{figure}
\end{figure}
In this subsection we will study the setting where the effective interaction is given by the beam splitter Hamiltonian (\ref{BShamiltonian}). This interaction leads to a coherent exchange of quantum states between the two mechanical modes at a rate $J_\mathrm{BS}$. Applications include the cooling of a low-frequency mode by means of an auxiliary high-frequency mode, the study of mediated interactions~\cite{mediatedinteractions} or quantum state mapping. In the time-domain the resonant coupling corresponds to a modulation of the strength of a spring connecting the two oscillators. It is instructive to discuss this condition in the frequency domain. 
\begin{figure}
\includegraphics[width=0.45\textwidth]{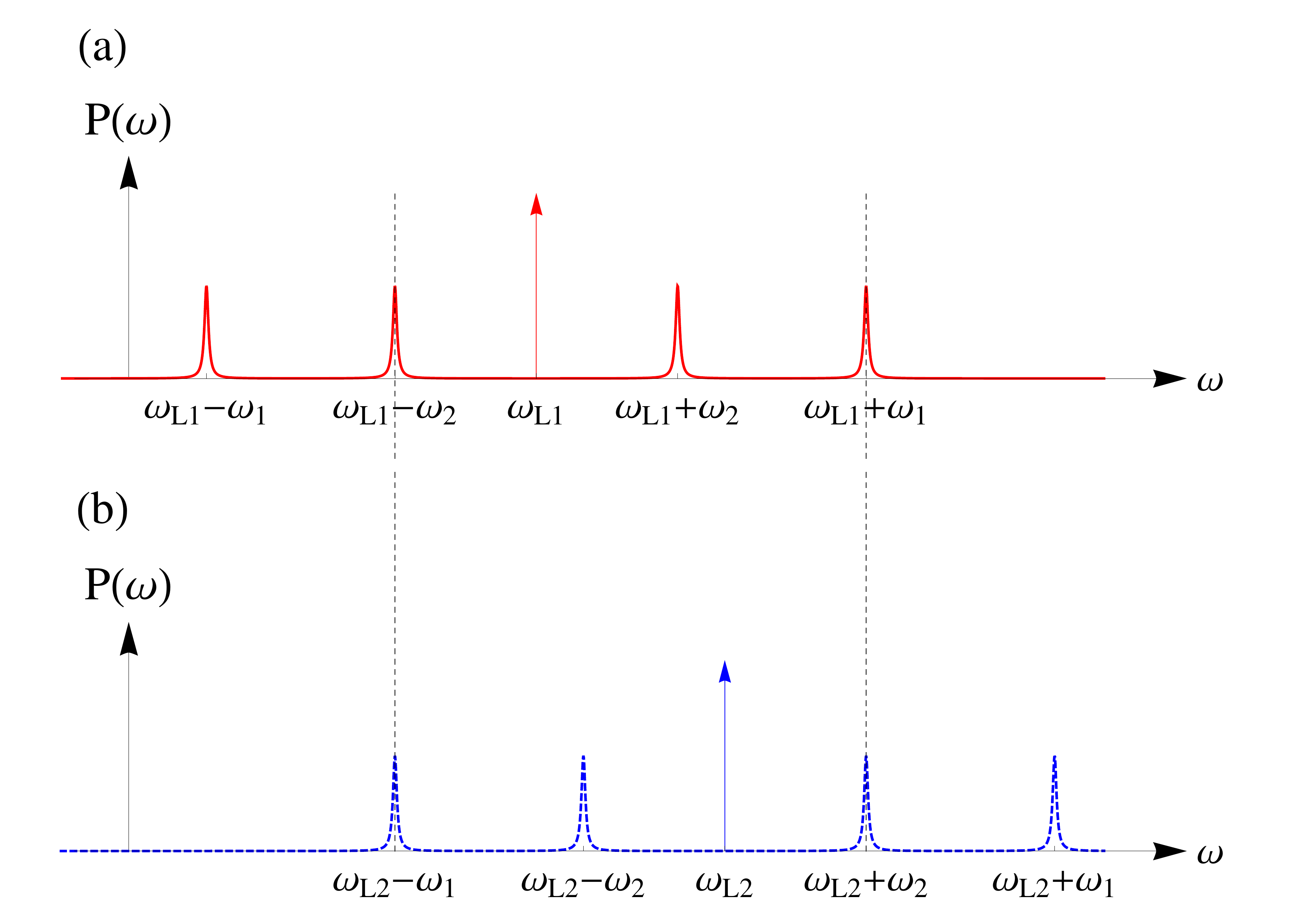}
\caption{(Color online) Sketch of the intracavity photon power spectral density neglecting the cavity profile for clarity. (a) and (b) show the contributions of the two pumps in red (solid) and blue (dashed) respectively.  The difference between the pump frequencies is arranged in the beam splitter configuration specified by Eq. (\ref{resonancecondition1}) with the pumps indicated by arrows representing delta-peaks at $\omega_{L1}$ and $\omega_{L2}$. Each pump creates a pair of sidebands for each mechanical oscillator, as indicated in the sketch. The overlapping sidebands are emphasized by the dashed lines and formally captured by Eqs. (\ref{matchedSBBS}). The exact behavior of this quantity including the cavity profile can be obtained from Eq. (\ref{coherences})}
\label{sidebandsketchBS}
\end{figure}
The optomechanical interaction leads to side bands on the cavity field, identifiable in Eq. (\ref{coherences}) and associated with photon-phonon scattering processes. Each pump creates two pairs of side bands, one for each oscillator. Their optical frequencies are
\begin{align}
\Delta_k\pm\omega_j. 
\end{align}
The resonance condition (\ref{resonancecondition1}) implies
\begin{subequations}\label{matchedSBBS}
\begin{align}
\Delta_1-\omega_1=\Delta_2-\omega_2=\bar\Delta-\bar\Omega\\
\Delta_1+\omega_2=\Delta_2+\omega_1=\bar\Delta+\bar\Omega,
\end{align}
\end{subequations}
where we have introduced the central detuning 
\begin{equation}
\bar\Delta=\frac{\Delta_{1}+\Delta_{2}}{2}
\end{equation}
These identities represent overlapping side bands of different oscillators, as sketched in Fig.~\ref{sidebandsketchBS}. 
 When a photon coming from one pump mode leaves the cavity in the other pump mode, it will have interacted with both oscillators and mediated the coherent transfer of a phonon between oscillators. The distinct overlapping side bands in Eqs.~(\ref{matchedSBBS}) provide two distinguishable pathways for phonon exchange and the coherent effective coupling rate is the sum of the amplitudes for each pathway. Additionally, there are four unmatched side bands with frequencies  
\begin{subequations}\label{unmatchedSBBS}
\begin{align}
\Delta_1+\omega_1,\\
\Delta_2-\omega_1,\\
\Delta_2+\omega_2,\\
\Delta_1-\omega_2. 
\end{align}
\end{subequations}
These unmatched side bands contain photons which have scattered phonons off a single mechanical mode. They carry information about the oscillators out of the cavity and are sources of dissipation and back-action. 
\par
The resonant coupling terms are found in the matrices $O_i^\pm$; using identities (\ref{matchedSBBS}) we find 
\begin{align}\label{couplingterms}
\left[O_1^\pm\right]_{12}=\left[O_1^\pm\right]_{21}=\left[O_2^\pm\right]_{12}^*=\left[O_2^\pm\right]_{21}^*=\frac{-g_1g_2\alpha_1\alpha_2}{\frac{\kappa}{2}+ i(\bar\Delta\mp\bar\Omega)},
\end{align}
where the star denotes complex conjugation. The coherent processes are contained in the imaginary parts of these coupling terms,
\begin{equation}
J_\mathrm{BS}=\mathrm{Im}\left(\left[O_1^+\right]_{12}+\left[O_1^-\right]_{12}\right),
\end{equation}
where the two terms correspond to amplitudes for the scattering being mediated by the matched side bands on the red and blue side of the pumps respectively. The real parts of Eq.~(\ref{couplingterms}) together with resonant terms in $O_3$ describe the dissipative dynamics. We find
\begin{align}
\left[O_3^\pm\right]_{12}=\left[O_3^\pm\right]_{21}=\frac{g_1g_2\alpha_1\alpha_2\kappa}{\kappa^2/4+(\bar\Delta\pm\bar\Omega)^2}.
\end{align}
Explicitly separating unitary and dissipative terms yields after some algebra the master equation
\begin{align}
\frac{d\rho_\mathrm{m}}{dt}=&-i\left[\mathcal{H}_\mathrm{eff},\rho_\mathrm{m}\right]+\bar\Gamma\left((\bar{n}+1)\mathcal{L}(\hat{B})\rho+\bar{n}\mathcal{L}(\hat{B}^\dag)\rho\right)\nonumber\\
&+\sum_{j=1}^2\bar\Gamma_j\left((\bar{n}_j+1)\mathcal{L}(\hat{b}_j)\rho+\bar{n}_j\mathcal{L}(\hat{b}_j^\dag)\rho\right)\label{effmastereqcouple}
\end{align}
with the effective Hamiltonian
\begin{equation}
\mathcal{H}_\mathrm{eff}=J_\mathrm{BS}\left(\hat{b}_1^\dag\hat{b}_2+\hat{b}_1\hat{b}_2^\dag\right),
\end{equation}
and the collective mode
\begin{equation}\label{colmode}
\hat{B}=\sqrt{\frac{G_1}{G_2}}\hat{b}_1+\sqrt{\frac{G_2}{G_1}}\hat{b}_2,
\end{equation}
where we have assumed $\alpha_1=\alpha_2\equiv\alpha$ and introduced $G_j=g_j\alpha$ for notational brevity. The chosen normalization of the collective mode yields non-canonical commutation relations but simplifies the coefficients appearing in the master equation. They are
\begin{subequations}\label{BSmastereqcoeff}
\begin{align}
J_\mathrm{BS}
=&\mathrm{Im}\left(\frac{G_1G_2(\kappa+2i\bar\Delta)}{\kappa^2/4+\bar\Omega^2-\bar\Delta^2+i\kappa\bar\Delta}\right),\label{jbs}\\
\bar\Gamma_1=&\frac{4G_1^2\kappa\bar\Delta(\bar\Omega+\delta\omega)}{(\kappa^2/4+(\bar\Delta-\bar\Omega-\delta\omega)^2)(\kappa^2/4+(\bar\Delta+\bar\Omega+\delta\omega)^2)},\label{gamma1barBS}\\
\bar\Gamma_2=&\frac{4G_2^2\kappa\bar\Delta(\bar\Omega-\delta\omega)}{(\kappa^2/4+(\bar\Delta+\bar\Omega-\delta\omega)^2)(\kappa^2/4+(\bar\Delta-\bar\Omega+\delta\omega)^2)},\label{gamma2barBS}\\
\bar\Gamma=&\frac{4G_1G_2\kappa\bar\Delta\bar\Omega}{(\kappa^2/4+(\bar\Delta-\bar\Omega)^2)(\kappa^2/4+(\bar\Delta+\bar\Omega)^2)},\label{gammabarBS}\\
\bar{n}_1=&\frac{\kappa^2/4+(\bar\Delta-\bar\Omega-\delta\omega)^2}{4\bar\Delta(\bar\Omega+\delta\omega)},\label{n1barBS}\\
\bar{n}_2=&\frac{\kappa^2/4+(\bar\Delta-\bar\Omega+\delta\omega)^2}{4\bar\Delta(\bar\Omega-\delta\omega)},\label{n2barBS}\\
\bar{n}=&\frac{\kappa^2/4+(\bar\Delta-\bar\Omega)^2}{4\bar\Delta\bar\Omega}.\label{nbarBS}
\end{align}
\end{subequations}
The master equation (\ref{effmastereqcouple}) describes two resonantly coupled bosonic modes connected to three independent heat baths. The two heat baths described by coupling rates $\bar\Gamma_j$ and effective thermal occupations $\bar{n}_j$ affect each oscillator separately, while the bath described by $\bar\Gamma$ with effective occupation $\bar{n}$ acts on a collective mode of the system. 
\subsubsection{Unitary Evolution}
\begin{figure}
\includegraphics[width=0.9\columnwidth]{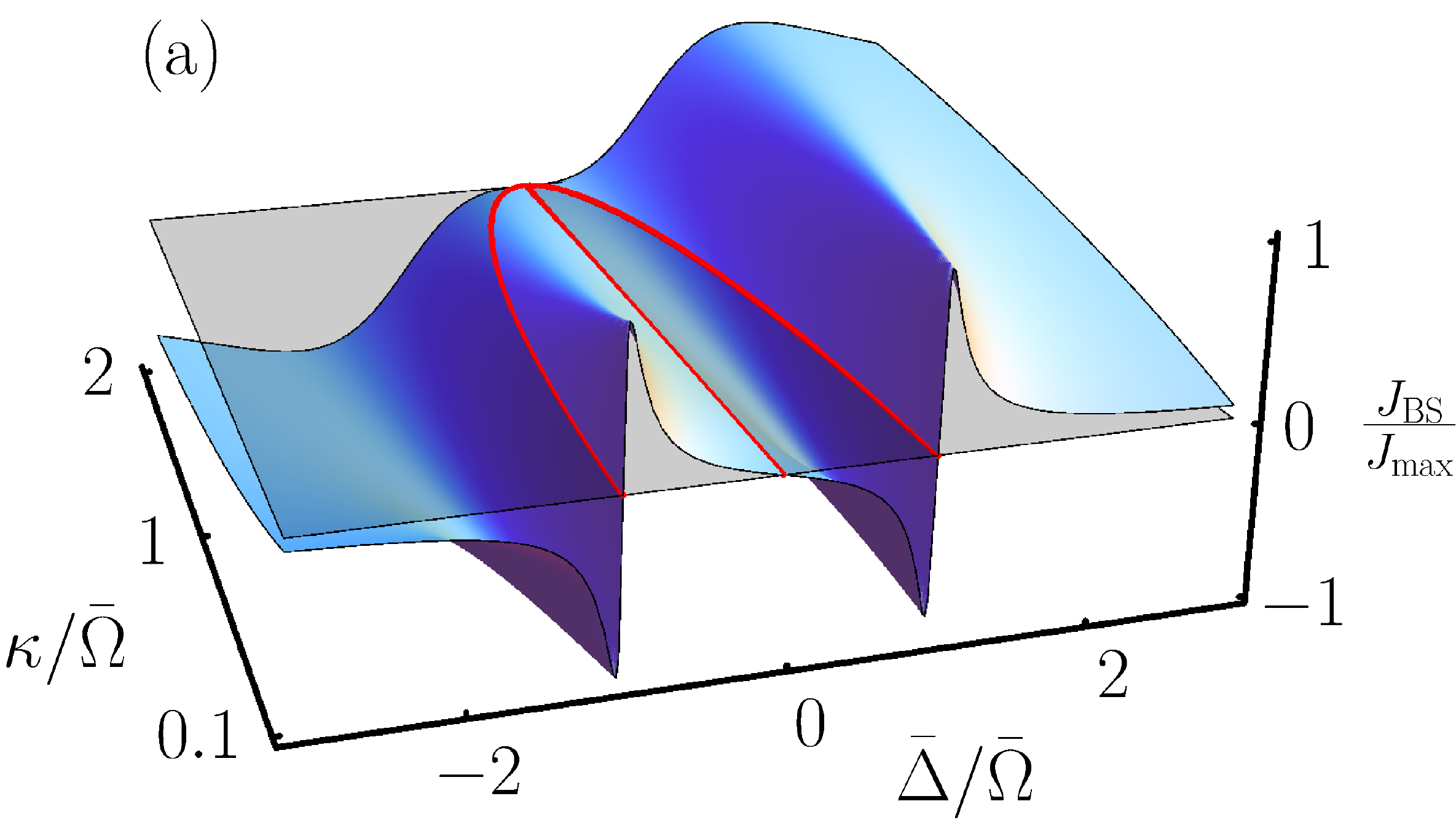}
\includegraphics[width=0.9\columnwidth]{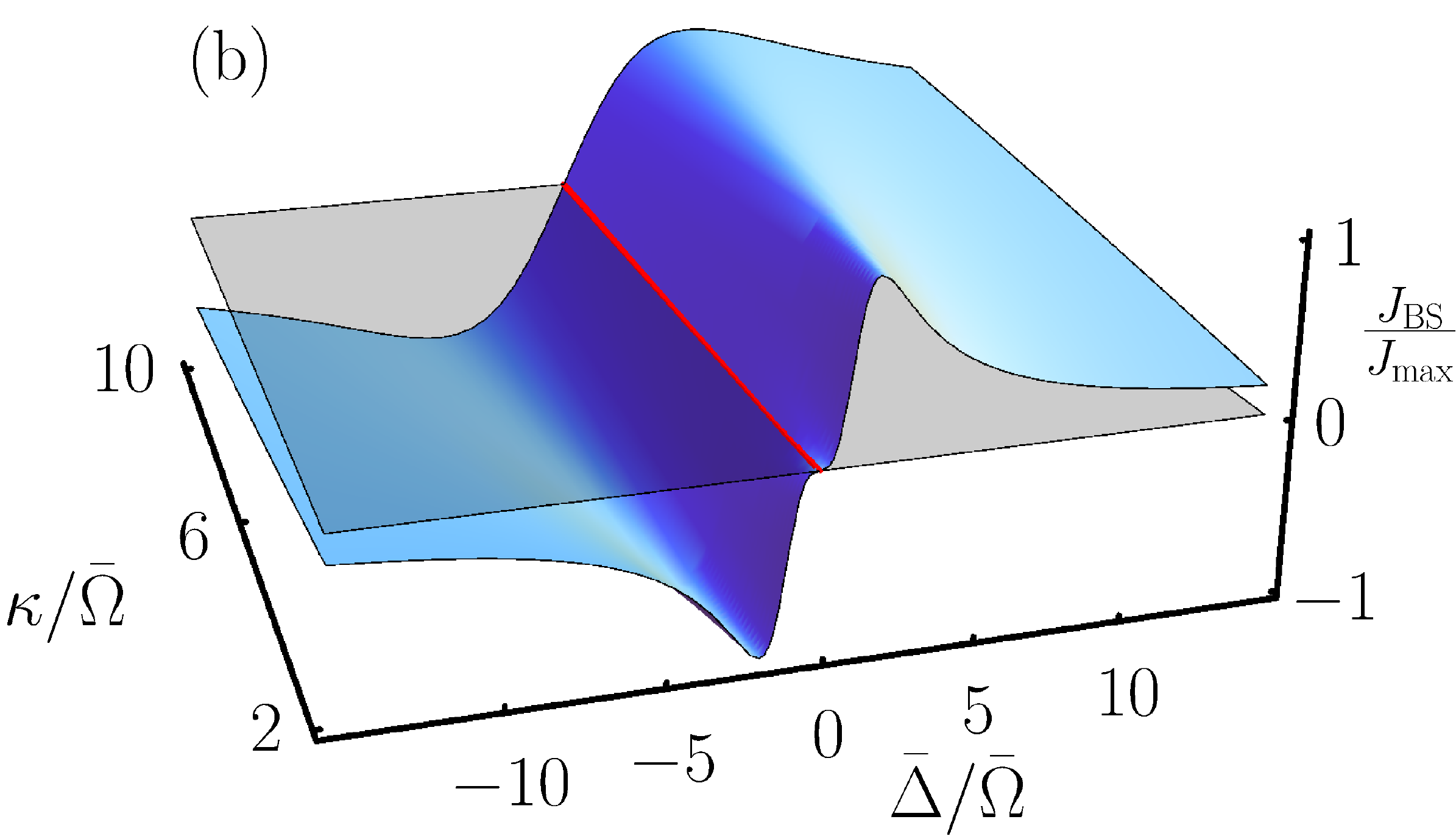}
\caption{(Color online) Normalized coupling strength as a function of central detuning and side band resolution. The line gives parameters resulting in zero coupling, $J_\mathrm{BS}=0$. Other parameters are $G_1=G_2=0.01~\bar\Omega$.}
\label{couplingfigures}
\end{figure}

For given side band resolution $\kappa/\bar{\Omega}$ and optomechanical couplings, the maximum exchange rate between the two states $J_\mathrm{max}$ is reached for \mbox{$\bar\Delta=\pm(\bar\Omega+\kappa/2)$}. For typical experimental parameters the coupling is around~$0.01~\bar\Omega$~\cite{review} and can be tuned via the pump strengths. The coupling strength as a function of central detuning changes its behavior depending on the side band resolution as plotted in Fig. \ref{couplingfigures}. 
For $\kappa<2\bar\Omega$ there are two non-zero values of $\bar\Delta$ resulting in no effective coupling, indicated by a line in the figures. This cancellation is a temporal analogue of a double-slit interference. The processes of phonon exchange via either matched side band (see Fig.~\ref{sidebandsketchBS} and Eqs. (\ref{matchedSBBS})) can coherently cancel. This is, however, only possible if the photon coherence time $\kappa^{-1}$ exceeds the inverse separation of the two processes $(\omega_1+\omega_2)^{-1}$. 

\subsubsection{Dissipative Dynamics}

The single-mode heat baths, characterized by Eqs. (\ref{gamma1barBS}),(\ref{gamma2barBS}),(\ref{n1barBS}) and (\ref{n2barBS}), correspond to the unmatched side bands of the two pumps, see Fig.~\ref{sidebandsketchBS}. These photons only carry information about the positions of single oscillators and affect the mechanical modes separately. Photons in the overlapping side bands partially contribute to the coherent exchange of excitations through $J_\mathrm{BS}$, but also leak out of the cavity, affecting the collective mode $\hat{B}$. 
Effective temperatures are defined by the occupations $\bar{n}_i$ and $\bar{n}$. Parameter regimes where $\bar{n}_i$ or $\bar{n}$ become negative are to be understood in the sense of statistical mechanics and indicate anti-damping of the mode in question.
\par
For pumps that are arranged symmetrically around the cavity resonance ($\bar\Delta=0$) there is no effective spring coupling the two oscillators, $J_\mathrm{BS}=0$. The light field only measures the mechanical system and its dissipation is pure measurement back-action of the light field on the mechanical oscillators~\footnote{The effective temperature diverges, but the quantity appearing in the master equation $\bar{\Gamma}\bar{n}$ remains finite}. Cavities with a dissipation rate smaller than the frequency difference of the oscillators can resolve each heat bath independently, as plotted in Fig.~\ref{decoherencefigure}. In such systems it is possible to cool or amplify each mode or the collective mode selectively and it is sufficient to consider only the dominant heat bath. 
\subsubsection{Limits to Coherence}
\begin{figure}
\includegraphics[width=0.9\columnwidth]{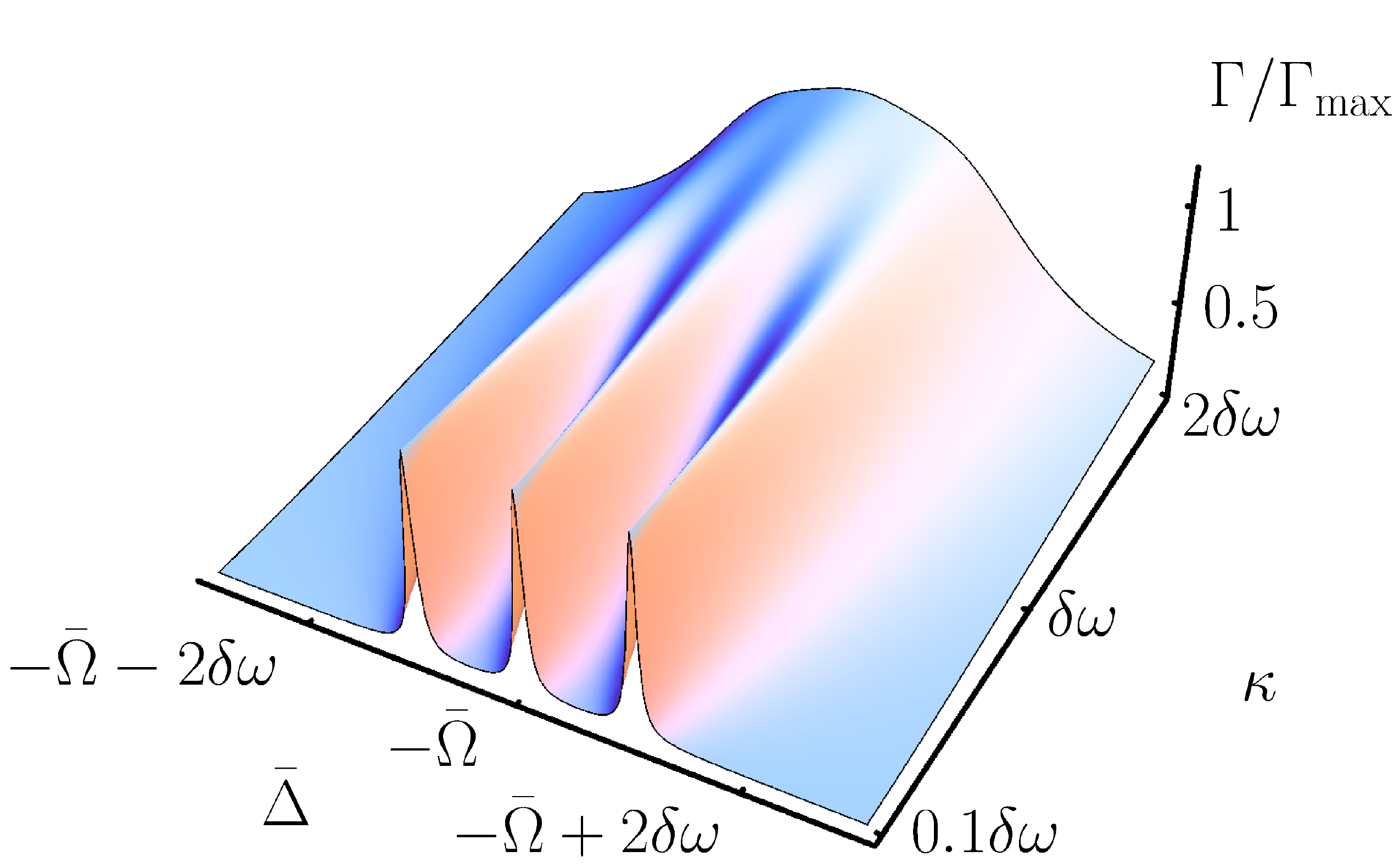}
\caption{(Color online) Scaled total decoherence rate $\Gamma/\Gamma_\mathrm{max}$ as a function of detuning and optical dissipation rate. Other parameters are the same as in Figs. \ref{couplingfigures}.}
\label{decoherencefigure}
\end{figure}

The detuning resulting in the strongest coupling strength $J_\mathrm{BS}$ is not necessarily the ideal choice to couple two mechanical modes. For coherent manipulations on a quantum level it is necessary to minimize the influence of decoherence. The total decoherence rate of the system due to the dissipation of the mediating field is given by
\begin{equation}
\Gamma=\bar\Gamma\bar{n}+\bar\Gamma_1\bar{n}_1+\bar\Gamma_2\bar{n}_2
\end{equation}
and is plotted in Fig. \ref{decoherencefigure} as a function of $\bar\Delta$ and $\kappa$. 
Each contribution is a Lorentzian with width $\kappa^2/4$ centered on the blue detuned side of the cavity resonance at $\bar\Delta=-\bar\Omega$ and $\bar\Delta=-\bar\Omega\pm\delta\omega$. 
For large $\kappa$, the ratio $J/\Gamma$ is typically smaller than unity for $\bar\Delta$ in the vicinity of $\bar\Omega$ and thus one would have to conclude that in this regime high-fidelity transfer of quantum states between oscillators is not possible. However, for $\bar\Delta\gg\bar\Omega$ this ratio becomes a monotonically increasing function of $\bar\Delta$ and can in principle be made arbitrary large~\footnote{The limiting factor is given by the available power in the pumps and the weak coupling regime.}. This remains true irrespective of the cavity's quality factor and is analogous to atomic physics, where the decoherence due to the spontaneous emission from an excited state can be circumvented by coupling two lower lying states via a far detuned two-photon transition. 
\subsection{Two-mode Parametric Amplification}
\begin{figure}
\includegraphics[width=0.45\textwidth]{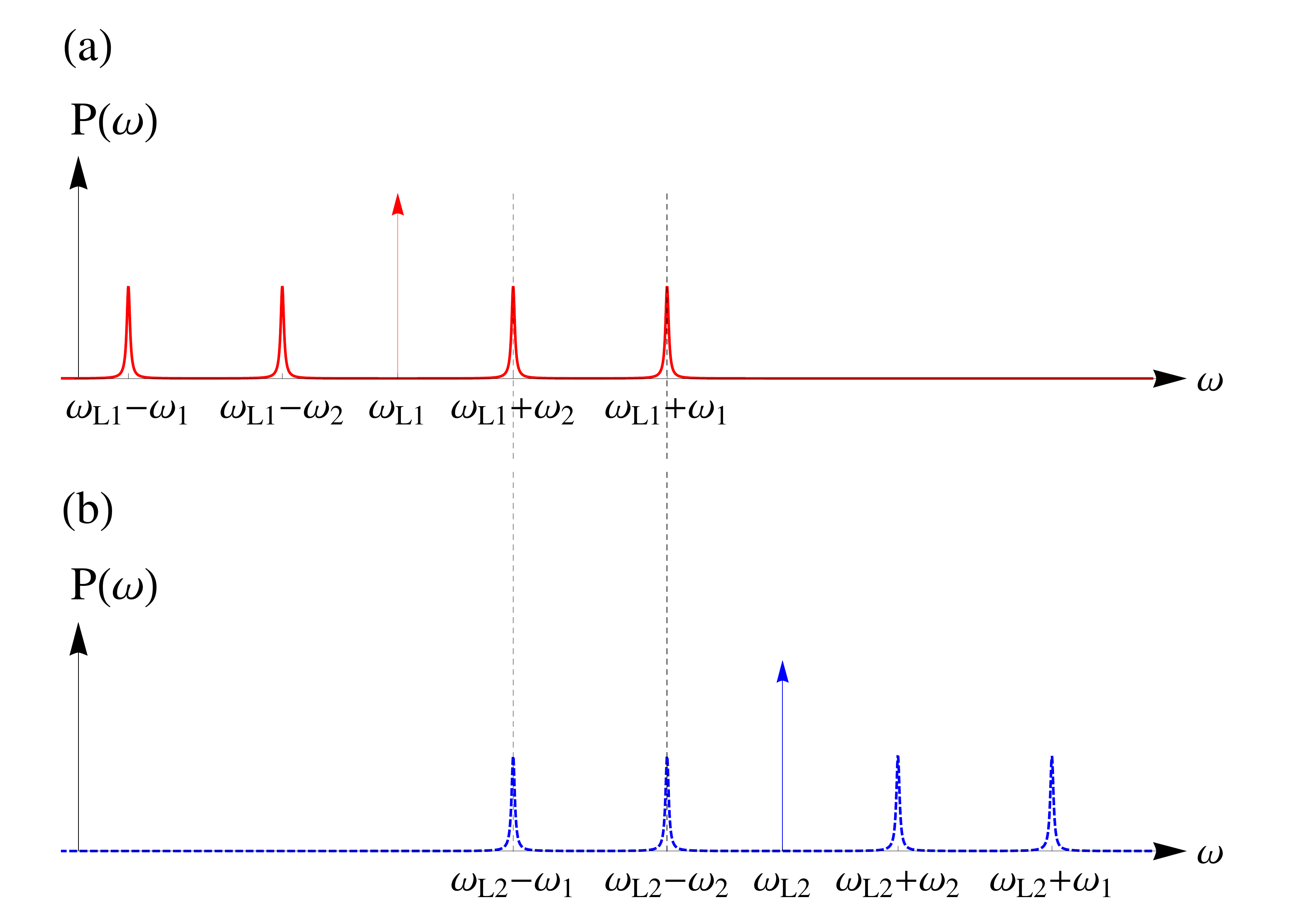}
\caption{(Color online) Sketch of the photon power spectral density neglecting the cavity profile. (a) and (b) show the power spectral density due to the two pumps in red (solid) and blue (dashed) respectively. The two pumps, indicated as arrows at $\omega_{L1}$ and $\omega_{L2}$, are arranged in the parametric amplifier configuration specified by Eq. (\ref{resonancecondition2}). We emphasized the overlapping sidebands captured by Eqs. (\ref{matchedSBPA}) with dashed lines.}\label{sidebandsketchPA}
\end{figure}

Let us now turn to the case where the resonant coupling results in two-mode parametric amplification as described by Hamiltonian (\ref{PAhamiltonian}). This interaction leads to fast growth of the amplitudes of both modes. The phase of each oscillator separately becomes uniformly distributed, while the phase of their collective motion becomes increasingly well defined. In the absence of dissipation, the resulting two-mode state is not separable and exhibits squeezing of the uncertainty of one collective quadratures of the system below its zero-point value~\cite{textbook}. 
Additionally this systems represents a pair of positive and negative mass oscillators containing a quantum-mechanics free subsystem \cite{negmassosc}. Even for degenerate modes, the interaction (\ref{PAhamiltonian}) needs auxiliary fields to become resonant. 
For the condition (\ref{resonancecondition2}) we easily find
\begin{subequations}\label{matchedSBPA}
\begin{align}
\Delta_1-\omega_2=\Delta_2+\omega_1=\bar{\Delta}+\delta\omega/2\\
\Delta_1-\omega_1=\Delta_2+\omega_2=\bar{\Delta}-\delta\omega/2.
\end{align}
\end{subequations}
A contributing resonant scattering process is given in Fig.~\ref{setupsketch}~(c) and the side band configuration in frequency space is sketched in panel (b) of Fig.~\ref{sidebandsketchPA}.
The pumps are arranged such that a blue side band of oscillator 1 is overlapped with the red side band of oscillator 2 and vice versa. Similar ``crossing of side bands'' was found in other set-ups to lead to entanglement \cite{huatang}, as is typically associated with parametric amplification.\\
The resonant coupling terms are found in the off-diagonal elements of the matrices $D_1^\pm$ and $D_2^\pm$. Using Eqs.~(\ref{matchedSBPA}), we find 
\begin{align}
\left[D_1^+\right]_{12}&=\left[D_1^-\right]_{21}=\left[D_2^+\right]_{12}^*=\left[D_2^-\right]_{21}^*\nonumber\\
&=\frac{-G_1G_2}{\kappa/2+i(\bar\Delta-\delta\omega/2)}\\
\left[D_1^+\right]_{21}&=\left[D_1^-\right]_{12}=\left[D_2^+\right]_{21}^*=\left[D_2^-\right]_{12}^*\nonumber\\
&=\frac{-G_1G_2}{\kappa/2+i(\bar\Delta+\delta\omega/2)}.
\end{align}
The resonant terms in the matrices $D_3^\pm$ contribute to dissipation
\begin{align}
\left[D_3^+\right]_{12}=\left[D_3^-\right]_{21}=\frac{G_1G_2\kappa}{\kappa^2/4+(\bar\Delta+\delta\omega)^2}\\
\left[D_3^+\right]_{21}=\left[D_3^-\right]_{12}=\frac{G_1G_2\kappa}{\kappa^2/4+(\bar\Delta-\delta\omega)^2}.
\end{align}
Writing out the master equation using these expressions and separating unitary from dissipative evolution, we arrive at 
\begin{align}\frac{d\rho_\mathrm{m}}{dt}=&-i\left[\mathcal{H}_{\mathrm{eff}},\rho_\mathrm{m}\right]+\bar\Gamma_-\mathcal{L}(\hat{B})\rho_\mathrm{m}+\bar\Gamma_+\mathcal{L}(\hat{B}^\dag)\rho_\mathrm{m}\nonumber\\
&+\sum_{j=1}^2\bar\Gamma_j\left((\bar{n}_j+1)\mathcal{L}(\hat{b}_j)\rho_\mathrm{m}+\bar{n}\mathcal{L}(\hat{b}_j^\dag)\rho_\mathrm{m}\right),\label{effmastereqPA}
\end{align}
where the effective Hamiltonian is 
\begin{align}
\mathcal{H}_\mathrm{eff}=J_\mathrm{PA}\left(\hat{b}_1\hat{b}_2+\hat{b}_1^\dag\hat{b}_2^\dag\right),
\end{align}
with 
\begin{align}
J_\mathrm{PA}
=\mathrm{Im}\left(\frac{G_1G_2(\kappa+2i\bar\Delta)}{\kappa^2/4+\delta\omega^2/4-\bar\Delta^2+i\kappa\bar\Delta}\right),\label{JPA}
\end{align}
and the collective mode is now
\begin{align}
\hat{B}=\sqrt{\frac{G_1}{G_2}}\hat{b}_1+\sqrt{\frac{G_2}{G_1}}\hat{b}_2^\dag.
\end{align}
Again, the commutation relation of this mode depends on the relative coupling strengths and vanishes for $G_1=G_2$, indicating a quantum-mechanics free subsystem~\cite{negmassosc}. 
Since $\hat{B}$ involves annihilation and creation operators, one cannot unambiguously assign an effective temperature to its reservoir. Instead, we will discuss the dynamics resulting from these dissipative terms from a numerical simulation later on. 
The parameters characterizing the dissipative evolution are 
\begin{subequations}
\begin{align}
\bar\Gamma_1&=\frac{4G_1^2\kappa\bar\Delta(2\bar\Omega+\delta\omega/2)}{\left[\frac{\kappa^2}{4}+(\bar\Delta-2\bar\Omega-\frac{\delta\omega}{2})^2\right]\left[\frac{\kappa^2}{4}+(\bar\Delta+2\bar\Omega+\frac{\delta\omega}{2})^2\right]},\label{gammabar1}\\
\bar\Gamma_2&=\frac{4G_2^2\kappa\bar\Delta(2\bar\Omega-\delta\omega/2)}{\left[\frac{\kappa^2}{4}+(\bar\Delta-2\bar\Omega+\frac{\delta\omega}{2})^2\right]\left[\frac{\kappa^2}{4}+(\bar\Delta+2\bar\Omega-\frac{\delta\omega}{2})^2\right]},\label{gammabar2}\\
\bar\Gamma_\pm&=\frac{G_1G_2\kappa}{\kappa^2/4+(\bar\Delta\pm\delta\omega/2)^2},\label{gammabarpm}\\
\bar{n}_1&=\frac{\kappa^2/4+(\bar\Delta-2\bar\Omega-\delta\omega/2)^2}{4\bar\Delta(2\bar\Omega+\delta\omega/2)},\label{nbar1}\\
\bar{n}_2&=\frac{\kappa^2/4+(\bar\Delta-2\bar\Omega+\delta\omega/2)^2}{4\bar\Delta(2\bar\Omega-\delta\omega/2)}.\label{nbar2}
\end{align}
\end{subequations}
 In contrast to the beam splitter case, the collective dynamics contained in Eqs. (\ref{JPA}) and (\ref{gammabarpm}) are independent of $\bar\Omega$ and depend on the frequency difference $\delta\omega$ only.

\subsubsection{Unitary Evolution}
\begin{figure}
\includegraphics[width=0.9\columnwidth]{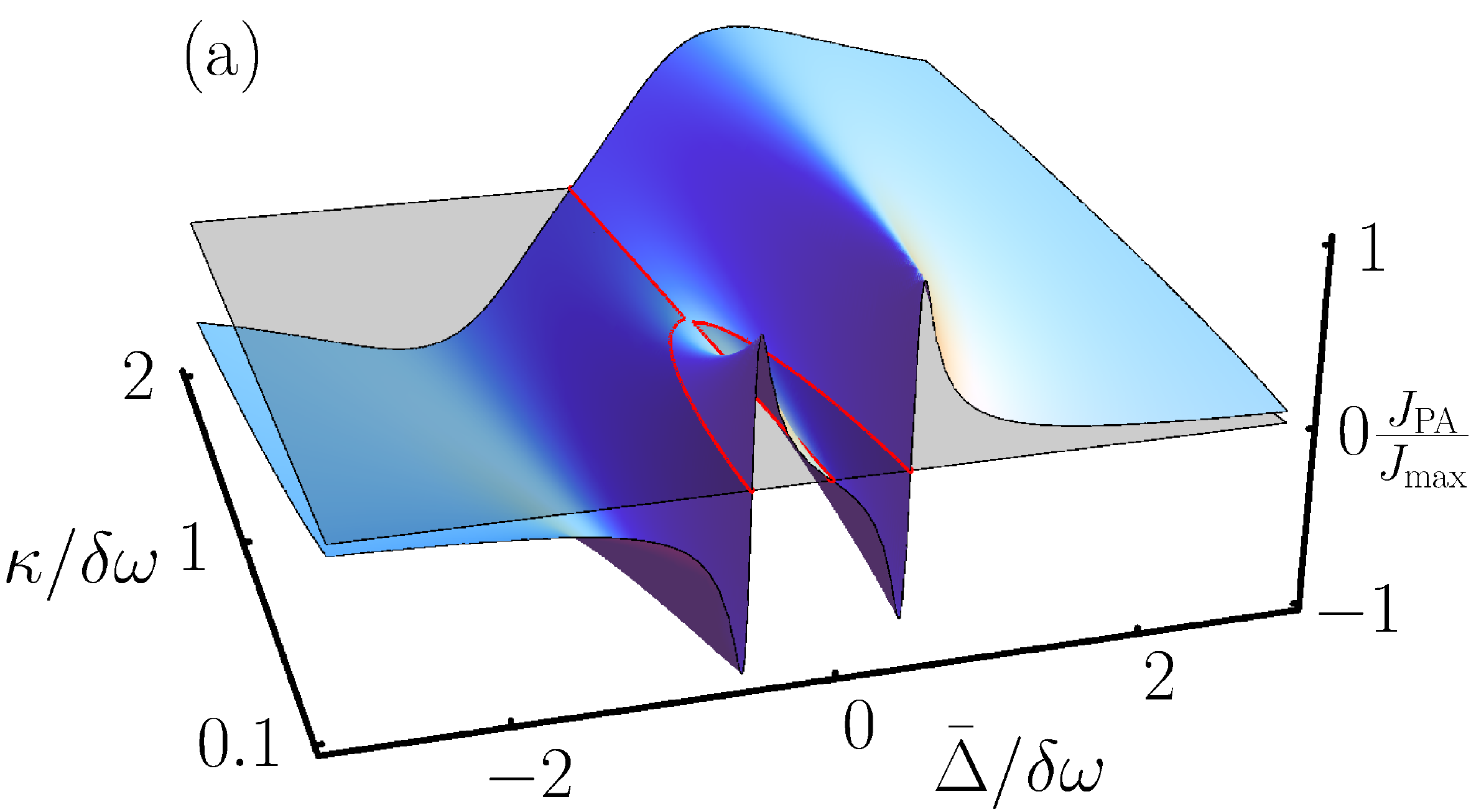}
\includegraphics[width=0.9\columnwidth]{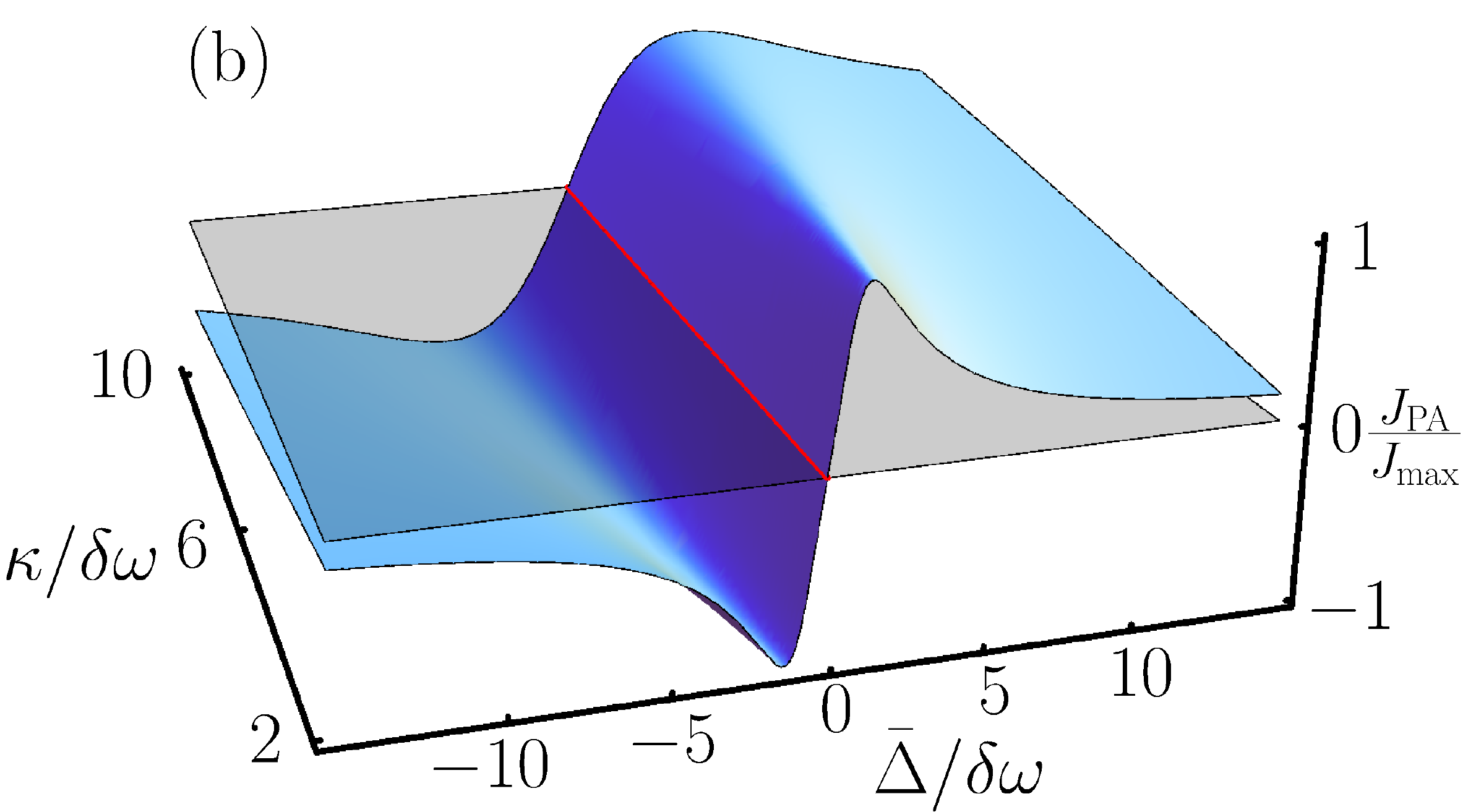}
\caption{(Color online) Normalized parametric driving strength as a function of central detuning and side band resolution. The line gives parameters with $J_\mathrm{PA}=0$. Other parameters are $G_1=G_2=0.01~\bar\Omega$. Note the formal equivalence up to a different normalization compared to Fig. \ref{couplingfigures}.}
\label{paramJfig}
\end{figure}

The parameter $J_\mathrm{PA}$ is proportional to the gain of the parametric amplifier and formally equivalent to $J_\mathrm{BS}$ with  the substitution $\bar\Omega\to\delta\omega/2$. It reaches its maximum $J_\mathrm{max}$ at $\bar\Delta=\pm\frac{1}{2}(\delta\omega+\kappa)$ and its qualitative behavior is plotted in Fig. \ref{paramJfig}. 
The single-quanta processes leading to parametric amplification can be found from Figs.~\ref{setupsketch}~(c) and \ref{sidebandsketchPA}. A photon is transferred from one pump to the other by the creation (or annihilation) of a phonon in each mechanical mode. This also clarifies the creation of entanglement by this interaction, as phonons in the two oscillators are created in pairs and the noise affecting each oscillator becomes correlated. Just as in the beam splitter configuration there are necessarily two distinct pathways for the coherent process separated in frequency by $\delta\omega$. Consequently systems with photon coherence times longer than the inverse of the frequency difference, $\kappa<\delta\omega$,  exhibit non-zero detunings where the pathways interfere destructively and lead to $J_\mathrm{PA}=0$. These detunings are drawn as lines in Fig.~\ref{paramJfig}.
\par
For non-vanishing $J_\mathrm{PA}$ the unitary interaction correlates the motion of both oscillators while increasing their energy exponentially. Whether the mediated interaction can entangle the two mechanical modes depends on the ratio of $J_\mathrm{PA}$ to the total decoherence rate in the system. 

\subsubsection{Dissipative Dynamics}
\begin{figure*}
\includegraphics[width=0.3\textwidth]{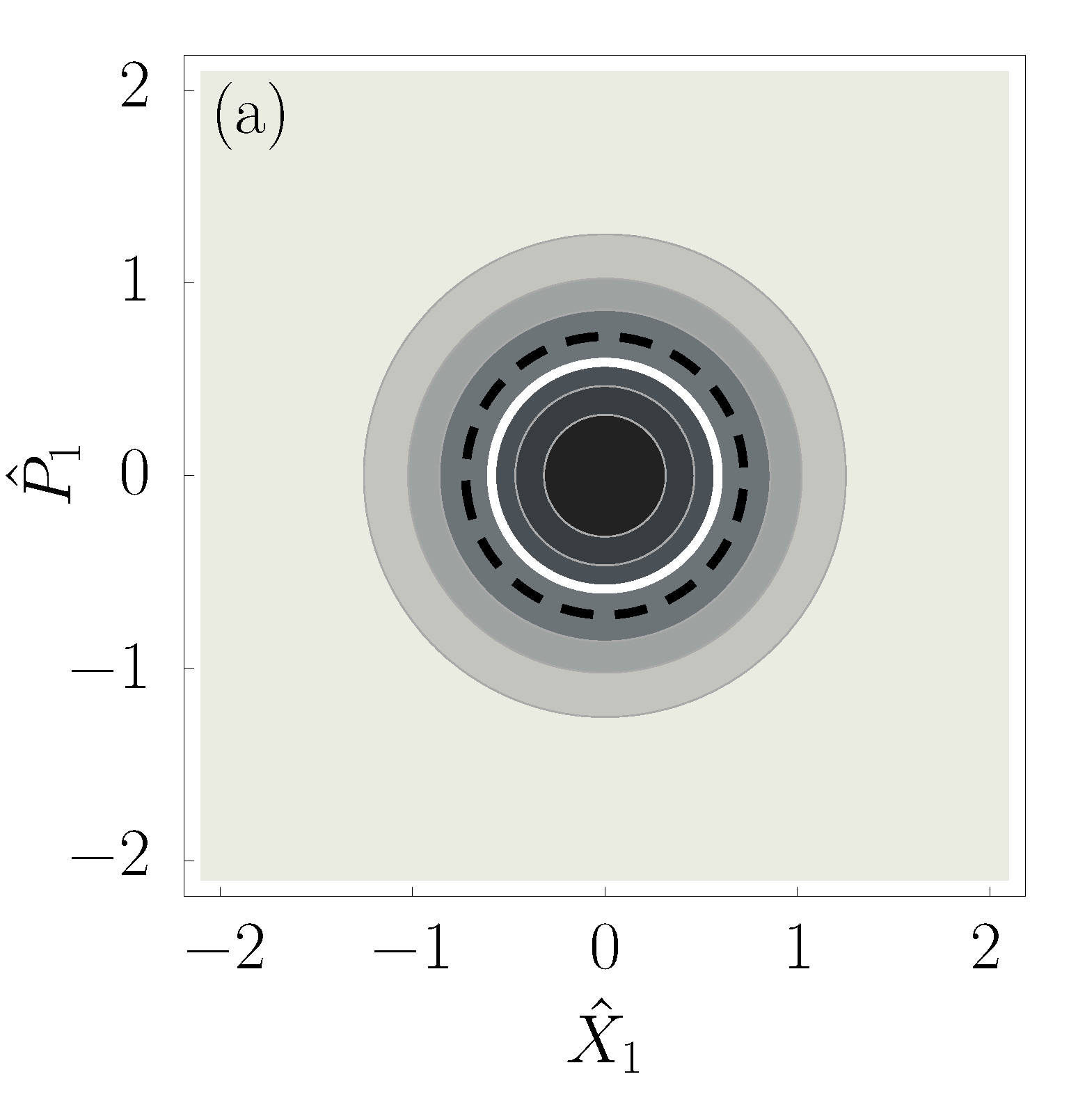}
\includegraphics[width=0.3\textwidth]{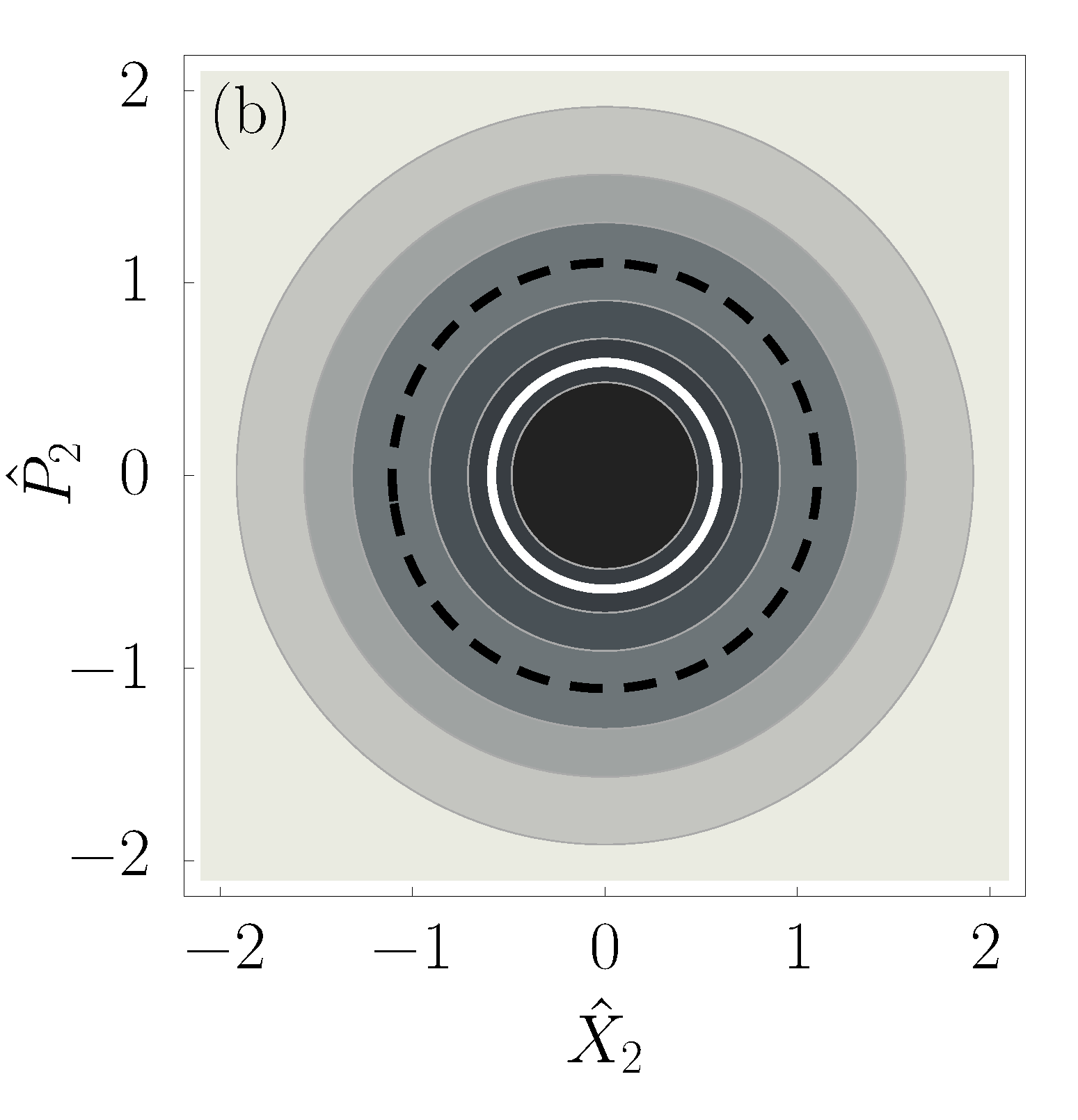}
\includegraphics[width=0.35\textwidth]{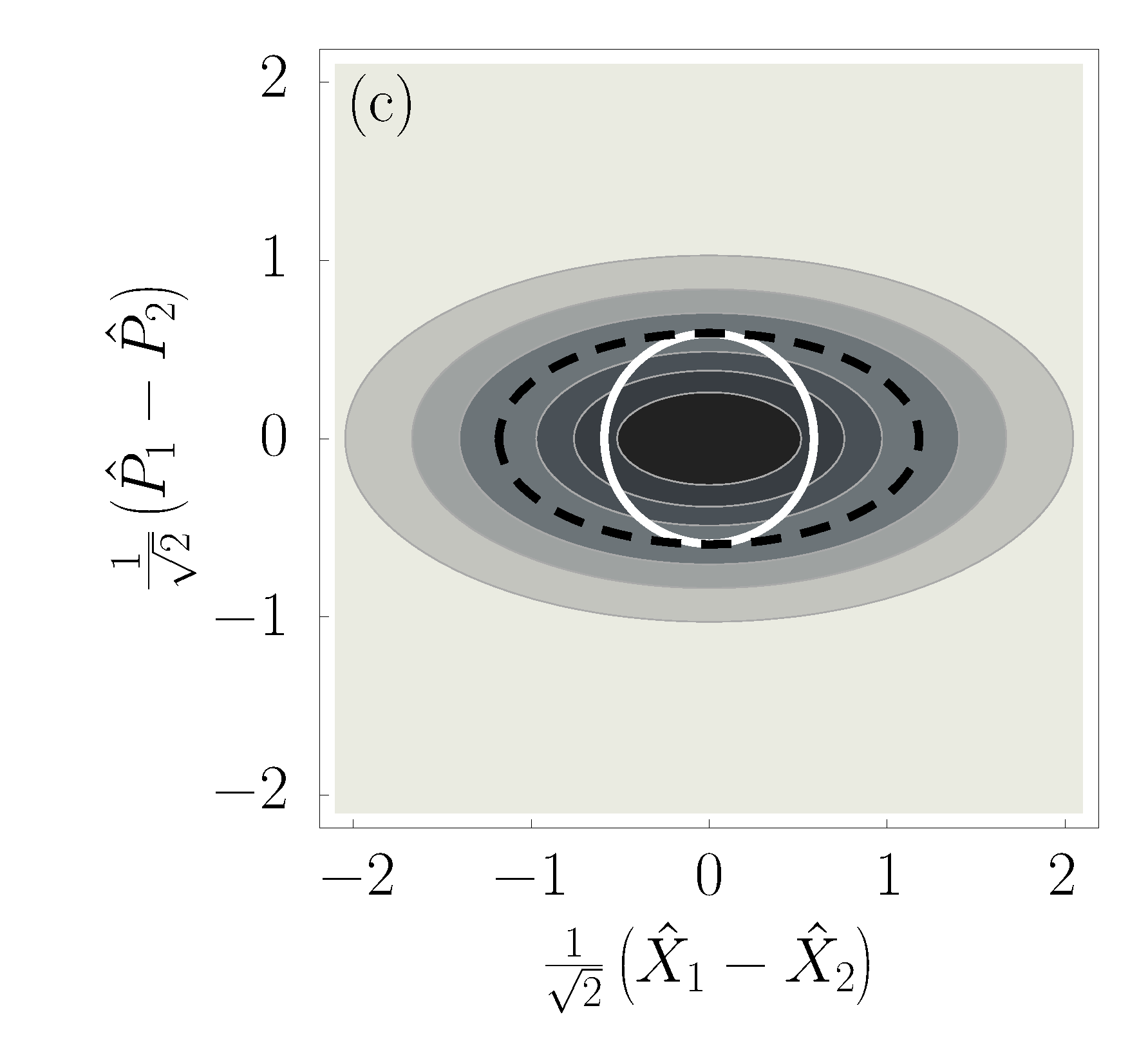}
\caption{(Color online) Phase-space probability distributions (Wigner functions) for: (a): oscillator 1, (b): oscillator 2, (c): collective mode under evolution of given by Eq.(\ref{dissterm}). The  dashed black line is the contour of half maximum, while the thick white one marks the same contour for a vacuum state.}
\label{probs} 
\end{figure*}

For the parametric amplifier the dissipative behavior is somewhat less intuitive than in the beam splitter situation. While the effect of the two pseudo-thermal, single-mode reservoirs leads to decoherence of each oscillator's quantum state with rate $\bar\Gamma_i(\bar n_i+1)$ towards a thermal state with occupations $\bar n_i$, the collective mode $\hat{B}$ is coupled to a non-trivial reservoir. 
To understand its phenomenology we solved the master equation 
\begin{equation}\label{dissterm}
\frac{d\rho_\mathrm{m}}{dt}=\Gamma_+\mathcal{L}(\hat{B})\rho_\mathrm{m}
\end{equation}
numerically for the initial vacuum state. The probability distributions for the two single mode oscillators and one of their collective modes after $t=\left(\Gamma_+\right)^{-1}$ are depicted in Figs. \ref{probs} together with half-maximum contours compared to their initial state. The dissipation term (\ref{dissterm}) increases the variance of both quadratures for each oscillator separately, albeit at a different rate. On the collective mode it heats only a  single quadrature, leaving the orthogonal one unchanged. These dynamics are the result of a back-action evading measurement on the collective mode $\hat{B}$. For $\bar\Delta\neq 0$, this evolution is accompanied by the diffusion of the orthogonal quadrature due to the non-vanishing value of $\Gamma_-$ in addition to the unitary evolution of parametric amplification. 

Just as in the beam splitter case, a cavity with small dissipation can resolve the different dissipation channels. If the cavity decay rate is smaller than the sum of the mechanical frequencies, $\kappa<2\bar\Omega$, one can choose whether the system couples to the single-mode reservoirs through (\ref{gammabar1}) and (\ref{gammabar2}) or to the collective, non-trivial reservoir (\ref{gammabarpm}). If the decay rate is furthermore smaller than the frequency difference, $\kappa<\delta\omega$, either one of the two different collective dynamics governed by Eq. (\ref{gammabarpm}) can be chosen separately.
\subsubsection{Entanglement}
\begin{figure}
\includegraphics[width=0.2\textwidth]{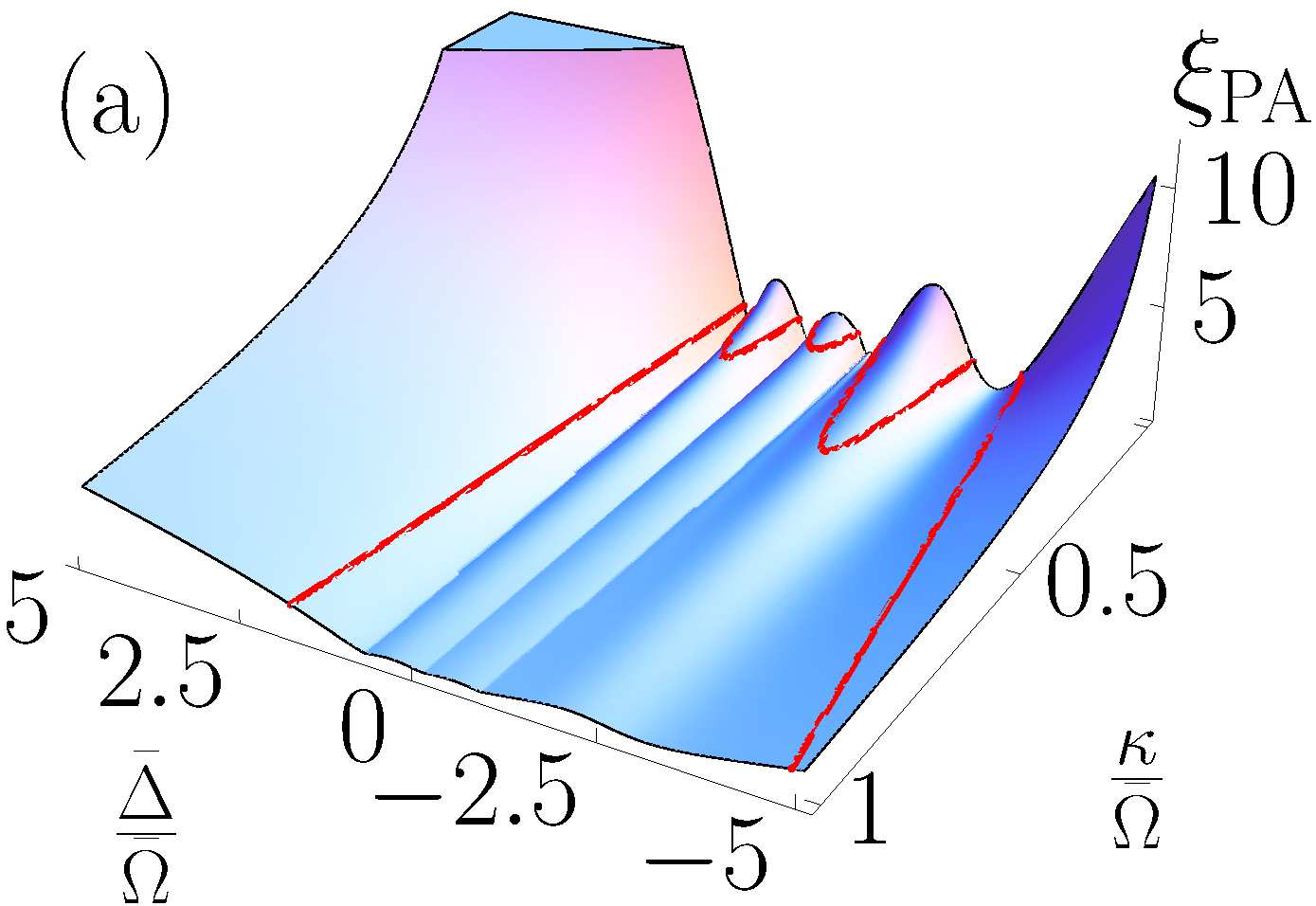}
\includegraphics[width=0.2\textwidth]{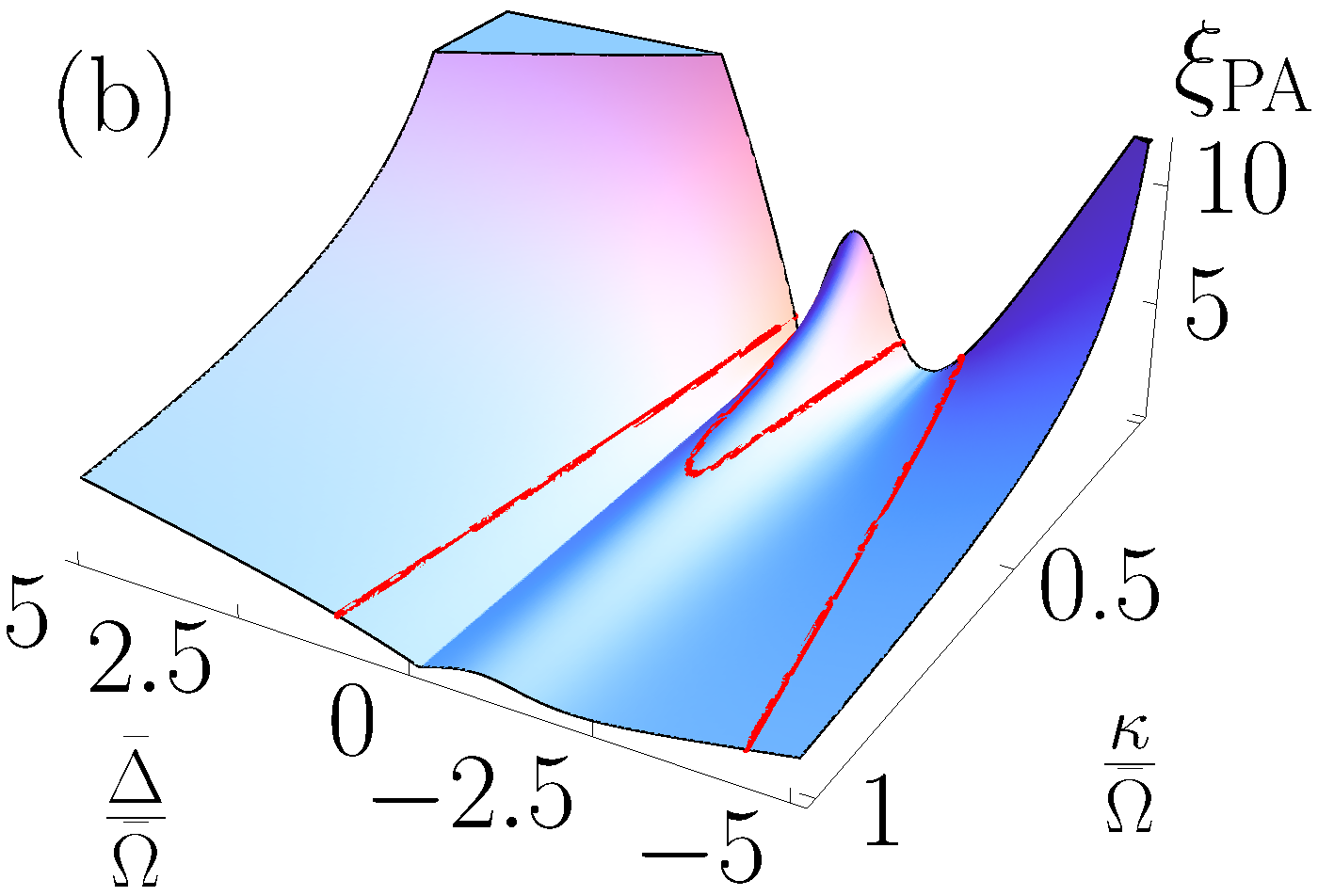}
\\
\includegraphics[width=0.2\textwidth]{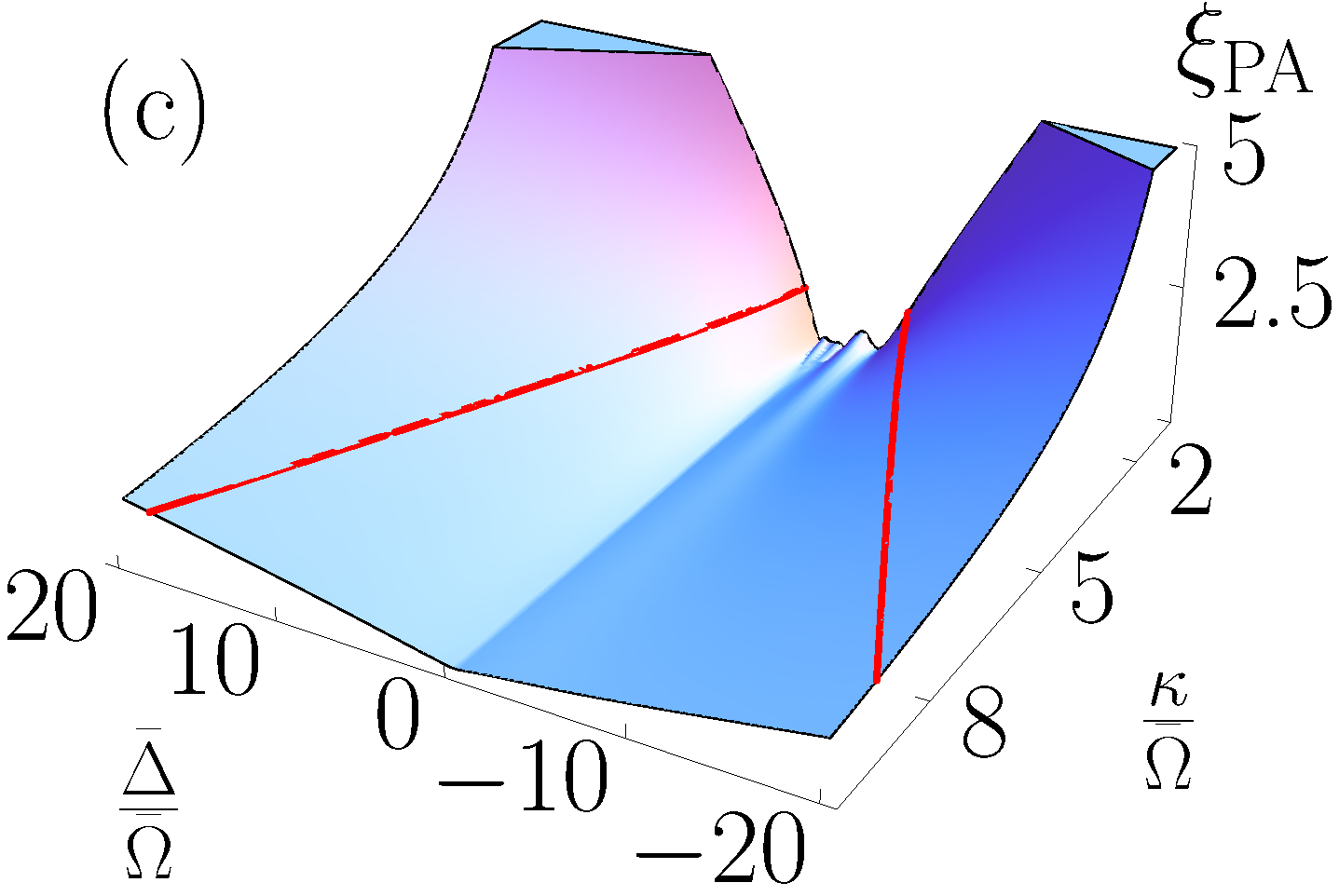}
\includegraphics[width=0.2\textwidth]{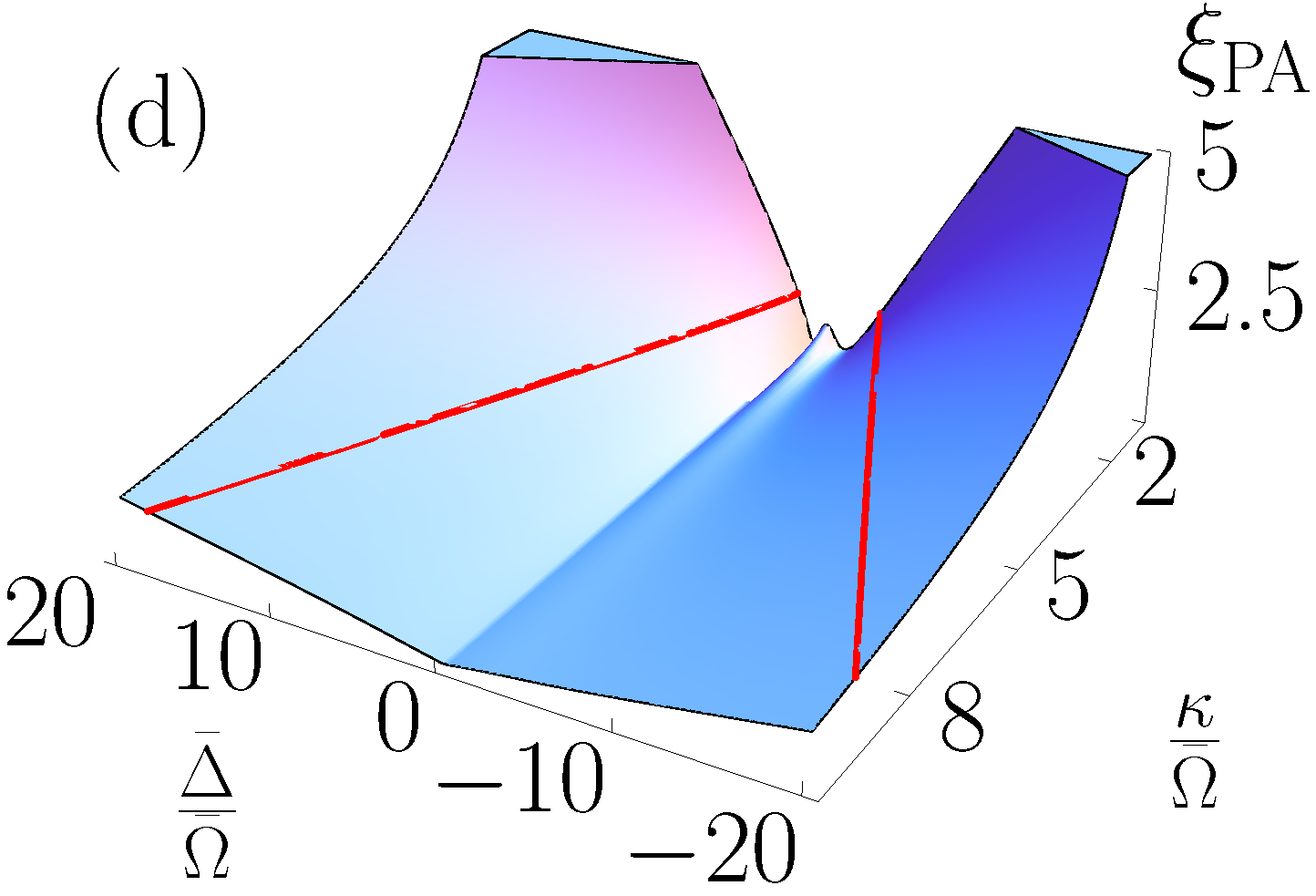}
\caption{(Color online) Effective coherences $\xi_\mathrm{PA}$ (as defined in Eq. (\ref{xipa})) as a function of central detuning $\bar\Delta$ and cavity decay rate $\kappa$.(a) and (c) are for widely separated oscillators, $\delta\omega/\bar\Omega=1.9$, while (b) and (d) are for $\delta\omega/\bar\Omega=0.1$. The red line denotes the value $\xi_\mathrm{PA}=1$. Other parameters are $G_1=G_2=0.05~\bar\Omega$.}
\label{xipafigs}
\end{figure}

The orientation of the squeezed collective quadrature in phase-space is not aligned with the squeezing from two-mode parametric amplification of the effective Hamiltonian (\ref{PAhamiltonian}), which occurs at an angle of $\pi/4$ with respect to the axes chosen in Fig. \ref{probs} (c)\cite{textbook}. In general this dissipative behavior will therefore counteract the entanglement generated by the unitary evolution. The relevant quantity which determines if the two oscillators can be driven into an inseparable state is the ratio
\begin{equation}\label{xipa}
\xi_\mathrm{PA}=\frac{J_\mathrm{PA}}{\bar\Gamma_1\bar n_1+\bar\Gamma_2\bar n_2+ \bar\Gamma_++\bar\Gamma_-}.
\end{equation}
Two oscillators can be driven into an entangled state for $\xi_{PA}>1$. We plotted this quantity in dependence of $\kappa$ and $\bar\Delta$ in Figs. \ref{xipafigs} for different parameter regimes. 
\par
The behavior of $\xi_\mathrm{PA}$ is qualitatively similar to the quantum/classical transition of linearly coupled oscillators found in~\cite{mediatedinteractions}. Perhaps most importantly, $\xi_\mathrm{PA}$ scales linearly with the central detuning $\bar\Delta$ for high optical decay rates $\kappa$ and can in principle be made arbitrarily large. For very large detunings this requires high pump powers, since the effective coupling decreases due to smaller intracavity photon numbers, $G_1G_2\propto 1/\bar\Delta^2$, and weaker effective coupling $J_{PA}\propto 1/\bar\Delta$, see Eq. (\ref{JPA}).
\par
Just as in the beam splitter case the ratio of unitary coupling strength to decoherence rate is analogous to two-photon transitions in atomic physics~\cite{textbook}. For large cavity-detunings the lossy photonic state becomes only virtually excited and thus decoherence from its dissipation can be made arbitrarily small. This allows to create non-classical correlations in a cavity with a low quality factor for sufficiently large central detunings and sufficiently high pump-powers.
\par
The dissipation rate in the denominator of Eq.~(\ref{xipa}) assumes that information contained in the photons emitted by the cavity is discarded. Using the outcomes of measurements performed on these photons to post-select data -- or alternatively using these photons directly to control a feed-back mechanism  -- would allow to create entangled mechanical modes even in regimes where $\xi_{PA}<1$. A detailed study of such mechanisms is beyond the scope of this work.

\subsection{Thermal Heat bath}
Throughout this article we have neglected the independent thermal heat baths for each oscillator. The derivation of the master-equation relies on the coupling strength of each oscillator to its bath $\gamma_i$ being much smaller than the cavity decay rate $\kappa$; a condition typically satisfied for optomechanical systems~\cite{review}.
The discussed coherent effects are negligible if thermal decoherence rates dominate the system, 
\begin{subequations}
\begin{align}\label{thermal1}
n_i\gamma_i\gg \frac{G_1G_2}{\kappa},
\end{align}
where $n_i$ denotes the occupation of oscillator $i$ths thermal reservoir. Even for such thermal systems, however, the resonant coupling can significantly alter the dynamics if the effective coupling is at least of the order of the reservoir couplings,
\begin{align}\label{thermal2}
\gamma_i\simeq \frac{G_1G_2}{\kappa}.
\end{align}
\end{subequations}
These conditions are not necessarily restrictive because each $G_i$ can be tuned via the intracavity photon numbers $\alpha_i^2$. 

\section{Summary}

We investigated a scheme to couple non-degenerate mechanical modes  resonantly using optomechanical coupling to a single cavity mode. Applying a bi-chromatic pump leads to time-dependent optical springs that can tune couplings between non-degenerate modes into resonance. We derived the reduced master equation for the two oscillators containing frequency shifts, unitary coupling and non-unitary effects such as dissipation and back-action. It is generally applicable for optomechanical systems in the weak coupling regime, ranging from macroscopic oscillators to ultracold atomic systems. We applied the master equation to discuss two resonant cases in depth: a beam splitter and a parametric amplifier. While the former allows more general cooling schemes and quantum state transfer between different frequency regimes, the latter could realize entanglement of spatially separated mechanical modes. We showed that these two fundamental quantum operations can be executed coherently irrespective of the cavity quality factor. Additionally, both cases exhibit mechanical quantum interference for sufficiently low cavity decay rates. 
\par
The coupling scheme allows customized interactions among distinct elements in arrays consisting of more than two mechanical modes~\cite{array} and the derived master equation is easily generalized to include more modes. 
Beyond mechanical excitations, the presented theory can be adapted to describe a variety of hybrid quantum systems, for instance ensembles of spins interacting with a single cavity mode~\cite{spinoptodyn} or spatially separated nitrogen vacancy centers in diamond coupled by a high-frequency phonon mode~\cite{nvdiamonds}. 

\begin{acknowledgements}
This work was supported by the Air Force Office of Scientific Research, NSF and a grant from the Swiss National Science Foundation.
\end{acknowledgements}

\pagebreak
\begin{appendix}
\begin{figure*}[bht]
\begin{tabular}{lll}
\includegraphics[width=0.325\textwidth]{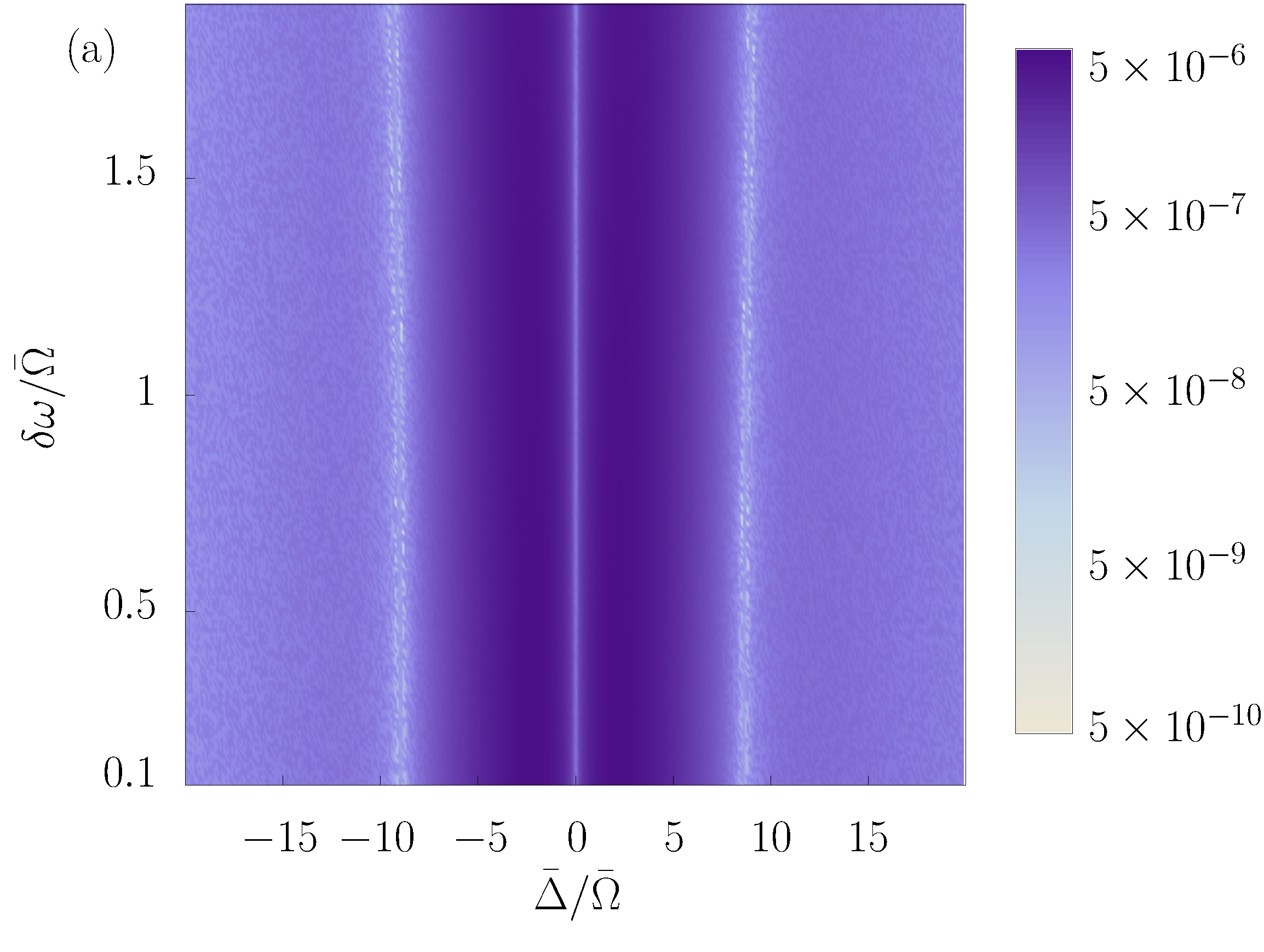}&
\includegraphics[width=0.30\textwidth]{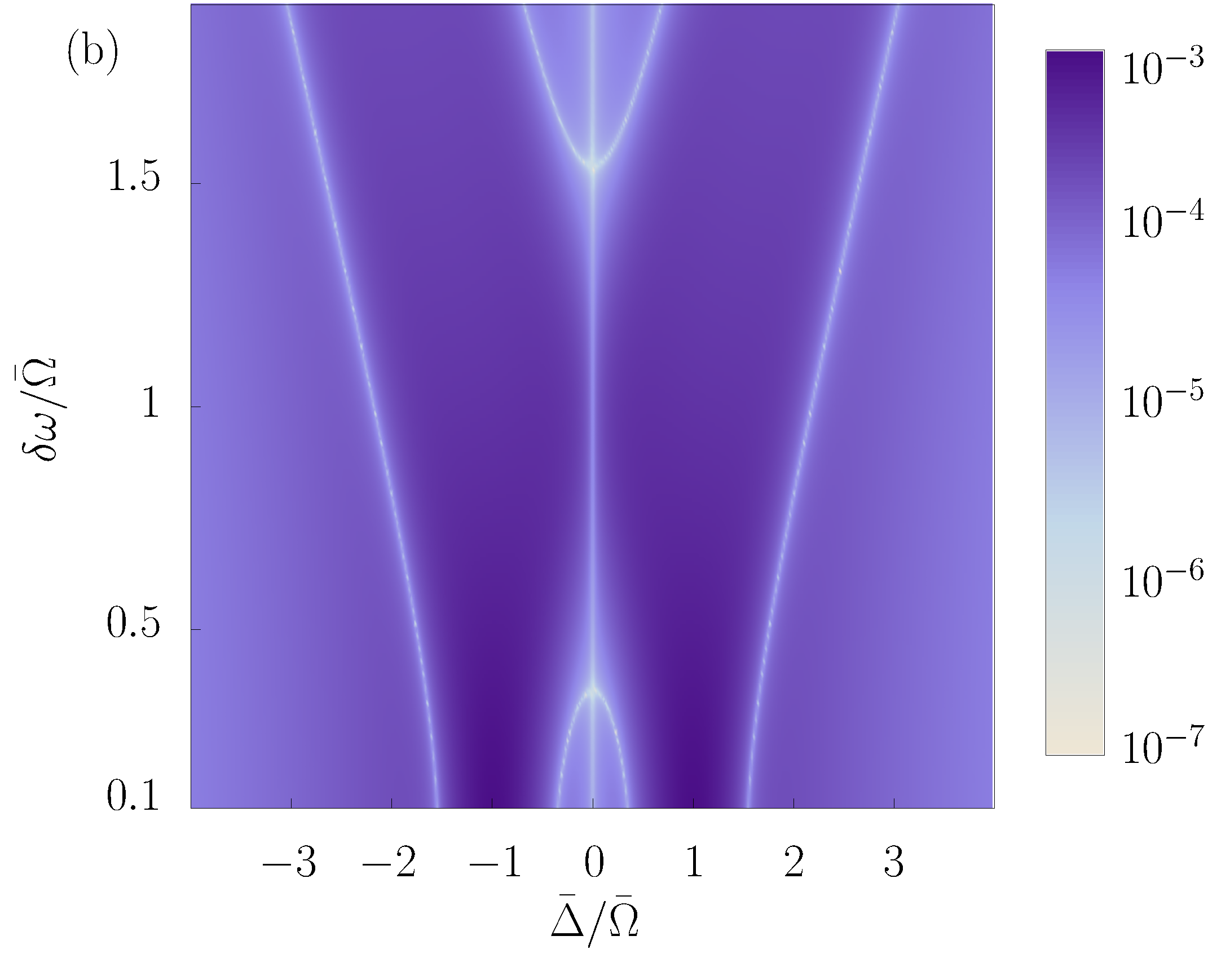}&
\includegraphics[width=0.315\textwidth]{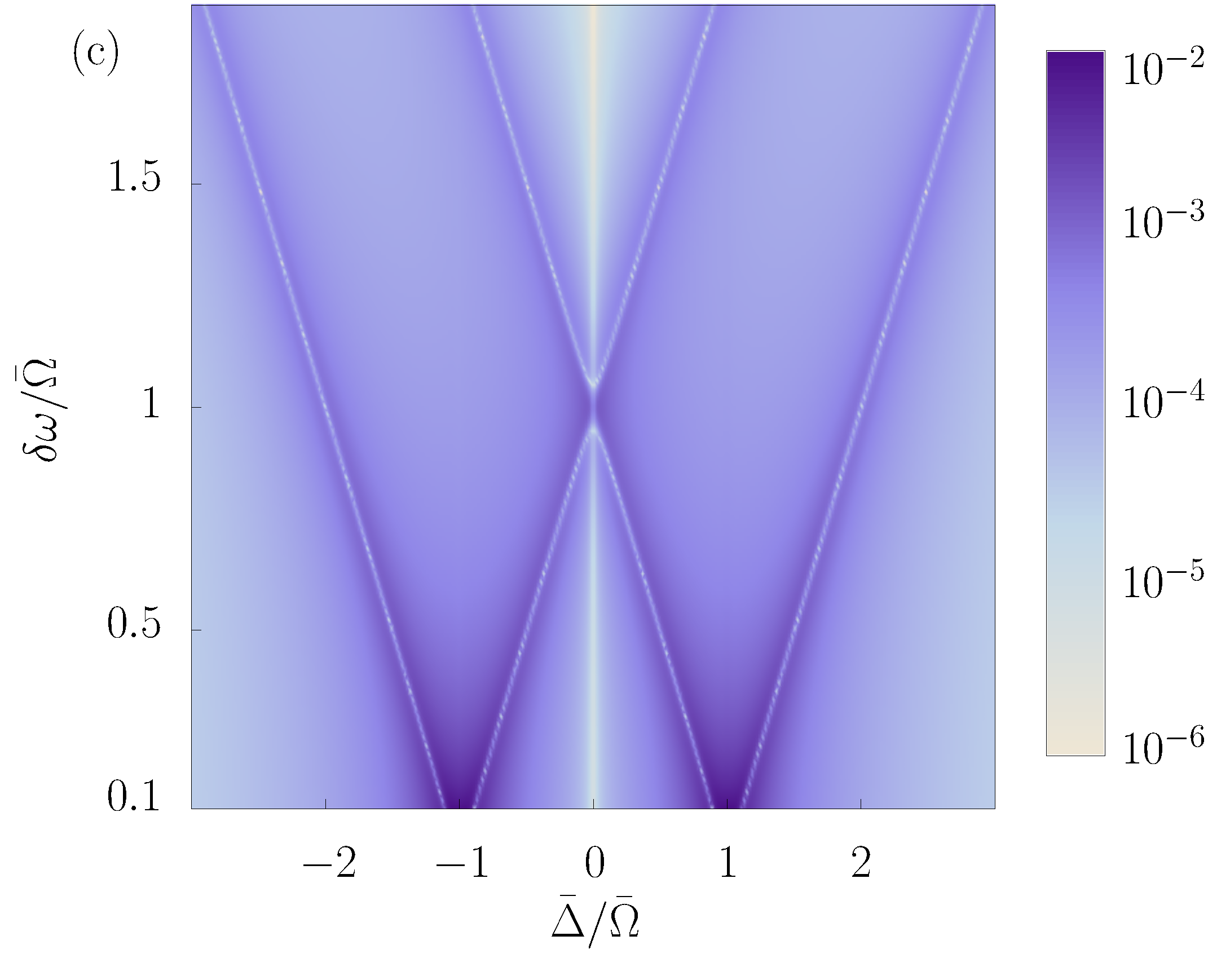}
\\
\includegraphics[width=0.305\textwidth]{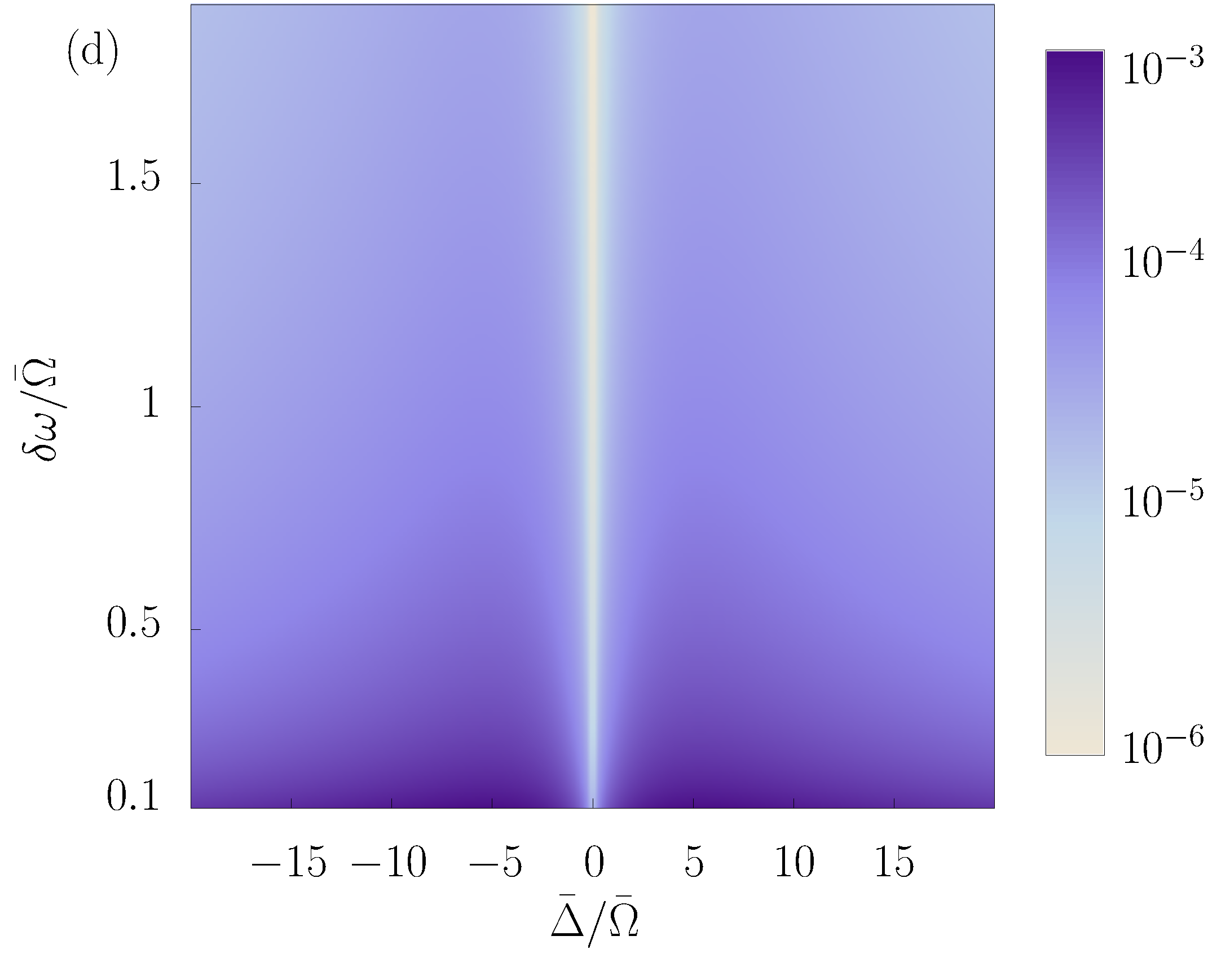}&
\includegraphics[width=0.325\textwidth]{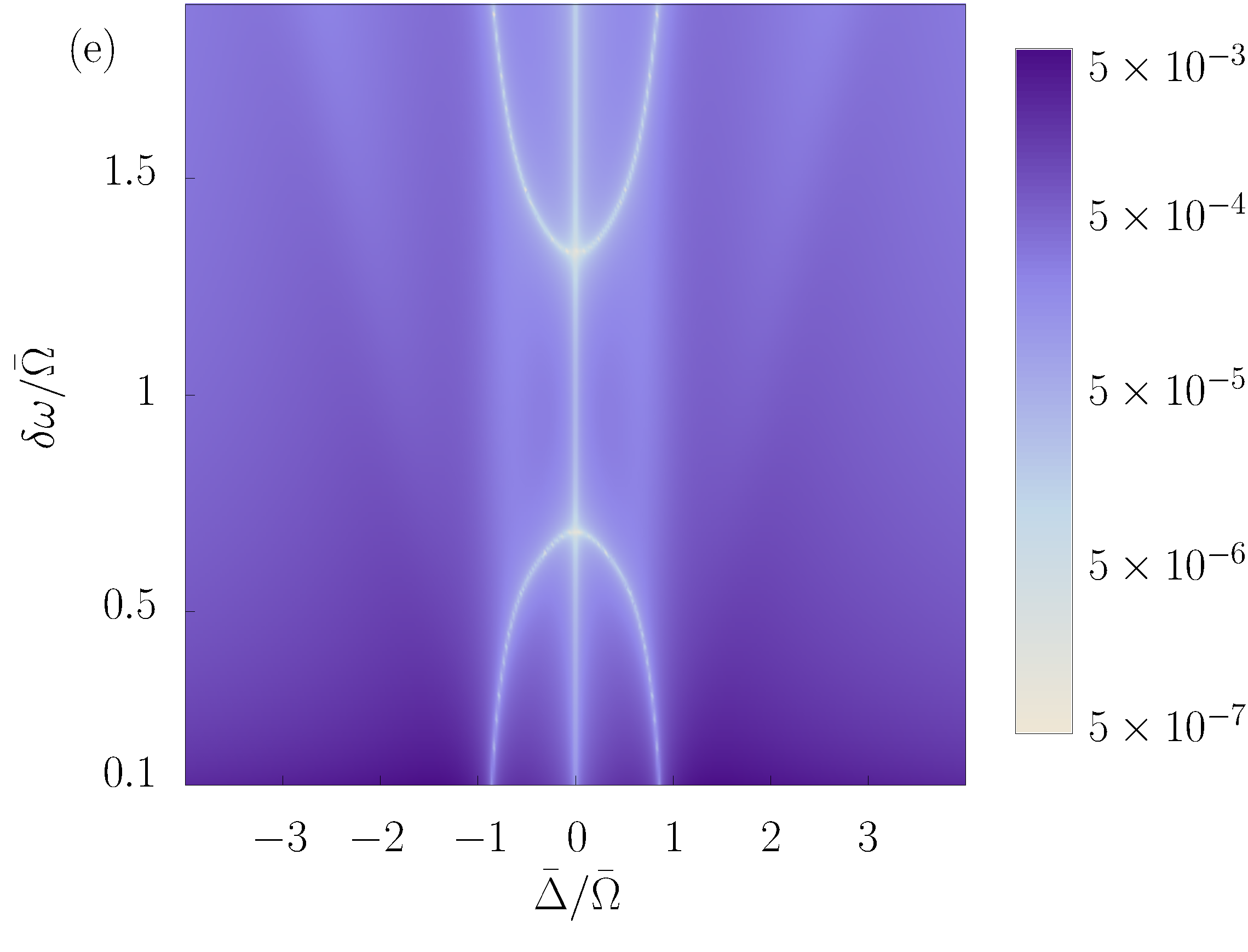}&
\includegraphics[width=0.315\textwidth]{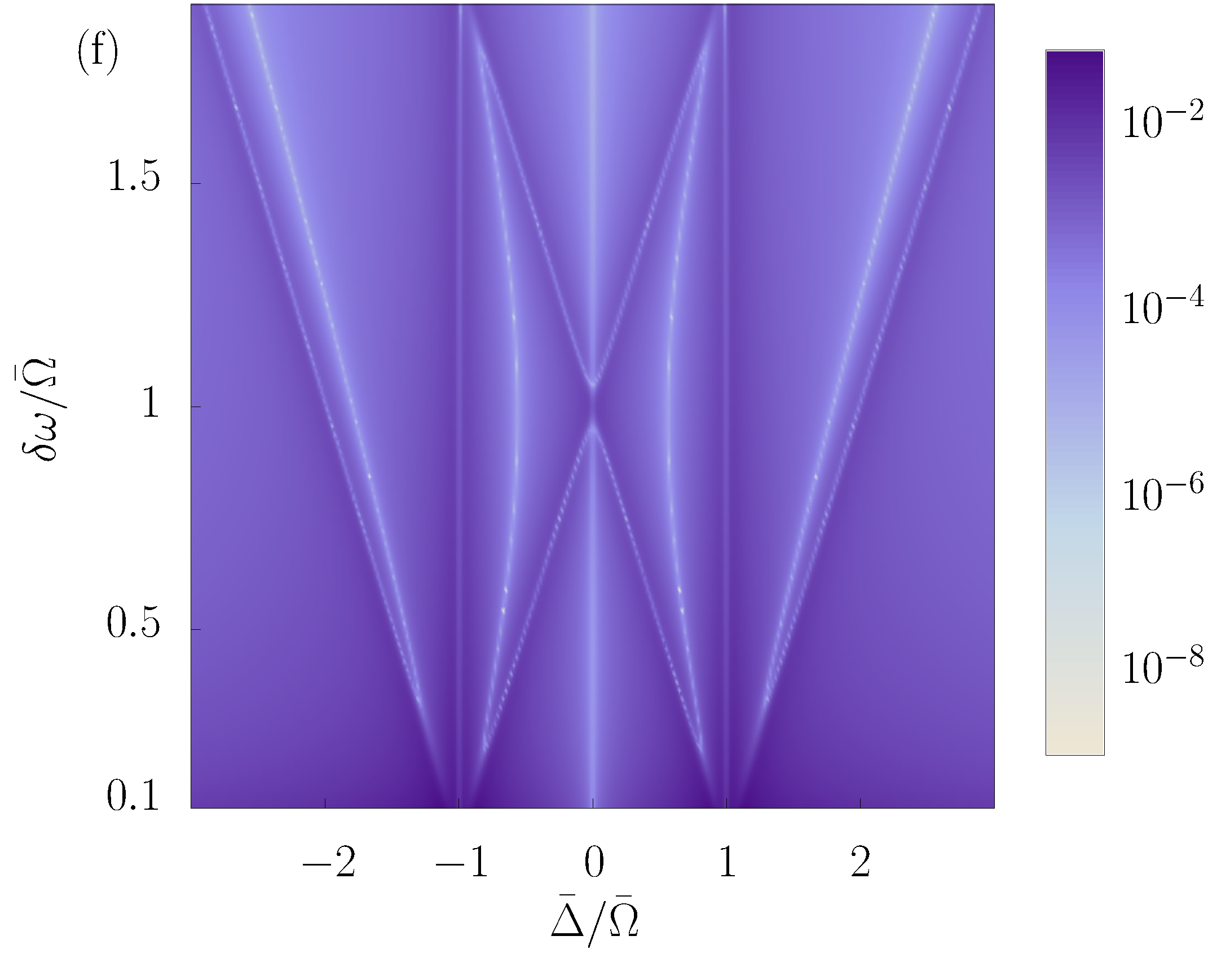}
\end{tabular}
\caption{(Color online) Corrections of the resonant pump detunings from their bare value. Top row for the beam splitter setting $\frac{\delta\Omega_1-\delta\Omega_2}{\delta\omega}$, bottom row for parametric amplification, $\frac{\delta\Omega_1+\delta\Omega_2}{2\bar\Omega}$. All plots have $g_j\alpha_k=0.01\bar\Omega$ and the optical dissipation rate is $\kappa=10\bar\Omega$ for (a) and (d), $\kappa=\bar\Omega$ for (b) and (e) and $\kappa=0.1\bar\Omega$ for (c) and (f).}\label{correctionplots}
\end{figure*}

\section{Nonlinear corrections to the resonance conditions}

Adopting the notation of Sec. \ref{mastereqsection}, with $\omega_i$ and $\delta\Omega_i$ denoting the bare mechanical frequencies and their optical spring shift respectively, the two modified resonance conditions read
\begin{align}
\Delta_1-\Delta_2=&\delta\omega+\delta\Omega_1-\delta\Omega_2 \label{modresonancecondition1}\\
\Delta_1-\Delta_2=&2\bar\Omega+\delta\Omega_1+\delta\Omega_2,\label{modresonancecondition2}.
\end{align}
These resonance conditions are nonlinear equations with respect to the two pump frequencies because the individual springs $\delta\Omega_i$ depend on the position of each pump with respect to the cavity resonance. While the individual shifts are subtracted from each other in the case of a beam splitter coupling, they add up for the parametric amplifier. 
For mechanical modes with identical couplings they remain small corrections to the bare frequency difference/sum, but since they provide limits to the fidelity of coherent state transfer or the gain of parametric amplification and the resulting entanglement, these effects have to be taken into account for applications in the quantum regime. The extent of the correction is measured by the dimensionless quantities 
\begin{align}
\frac{\delta\Omega_1-\delta\Omega_2}{\delta\omega}\\
\frac{\delta\Omega_1+\delta\Omega_2}{2\bar\Omega},
\end{align}
for the beam splitter and parametric amplifier respectively. We have calculated these expressions numerically and plotted them as a function of the bare mechanical frequency difference $\delta\omega$ and the average of the pump detunings  in Fig.~\ref{correctionplots}.

\end{appendix}

\end{document}